\documentclass[11pt,a4paper]{article}
\pdfoutput=1

\usepackage{jheppub}
\usepackage{latexsym}
\usepackage{color}
\usepackage[usenames,dvipsnames,svgnames,table]{xcolor}
\usepackage{graphicx}
\usepackage{epsfig}  
\usepackage{epsf}   
\usepackage{dcolumn}
\usepackage{bm}
\usepackage[toc,page]{appendix}
\usepackage{dcolumn}
\usepackage{textcomp}
\usepackage{float}
\usepackage{hypcap}
\usepackage[]{hyperref}
\usepackage{booktabs}
\usepackage{multirow}
\usepackage{diagbox}
\usepackage{alphalph}
\usepackage{textcomp}
\usepackage{epstopdf}
\usepackage{orcidlink}
\usepackage{url}

\usepackage[latin1]{inputenc} 
\usepackage[normalem]{ulem}

\newcommand{\capdef}{}
\newcommand{\mycaption}[2][\capdef]{\renewcommand{\capdef}{#2}
\caption[#1]{{\footnotesize #2}}}

\newcommand{\beq}{\begin{equation}}
\newcommand{\eeq}{\end{equation}}
\newcommand{\beqa}{\begin{eqnarray}}
\newcommand{\eeqa}{\end{eqnarray}}
\definecolor{orcidlogocol}{HTML}{A6CE39}

\preprint{IP/BBSR/2024-04}

\title{Improved precision on 2-3 oscillation parameters using the synergy between DUNE and T2HK}

\author[a,b,c]{\orcidlink{0000-0002-9714-8866}Sanjib Kumar Agarwalla,}
\author[a,b]{\orcidlink{0000-0003-3258-4357}Ritam Kundu,}
\author[a,d]{\orcidlink{0000-0002-8363-7693}Masoom Singh}

\affiliation[a]{Institute of Physics, Sachivalaya Marg, Sainik School Post, Bhubaneswar 751005, India}
\affiliation[b]{Homi Bhabha National Institute, Training School Complex, Anushakti Nagar, Mumbai 400094}
\affiliation[c]{Department of Physics \& Wisconsin IceCube Particle Astrophysics Center, University of Wisconsin, Madison, WI 53706, U.S.A}
\affiliation[d]{Department of Physics, Utkal University, Vani Vihar, Bhubaneswar 751004, India}

\emailAdd{sanjib@iopb.res.in}
\emailAdd{ritam.k@iopb.res.in}
\emailAdd{masoom@iopb.res.in}

\abstract{A high-precision measurement of $\Delta m^2_{31}$ and $\theta_{23}$ is inevitable to estimate the Earth's matter effect in long-baseline experiments which in turn plays an important role in addressing the issue of neutrino mass ordering and to measure the value of CP phase in $3\nu$ framework. After reviewing the results from the past and present experiments, and discussing the near-future sensitivities from the IceCube Upgrade and KM3NeT/ORCA, we study the expected improvements in the precision of 
2-3 oscillation parameters that the next-generation long-baseline experiments, DUNE and T2HK, can bring either in isolation or combination. We highlight the relevance of 
the possible complementarities between these two experiments in obtaining the improved sensitivities in determining the deviation from maximal mixing of $\theta_{23}$, excluding the wrong-octant solution of $\theta_{23}$, and obtaining high precision on 2-3 oscillation parameters, as compared to their individual performances. We observe that for the current best-fit values of the oscillation parameters and assuming normal mass ordering (NMO), DUNE + T2HK can establish the non-maximal $\theta_{23}$ and exclude the wrong octant solution of $\theta_{23}$ at around 7$\sigma$ C.L. with their nominal exposures. We find that DUNE + T2HK can improve the current relative 1$\sigma$ precision on $\sin^{2}\theta_{23}~(\Delta m^{2}_{31})$  by a factor of 7 (5) assuming NMO. Also, we notice that with less than half of their nominal exposures, the combination of DUNE and T2HK can achieve the sensitivities that are expected from these individual experiments using their full exposures. We also portray how the synergy between DUNE and T2HK can provide better constraints on ($\sin^2\theta_{23}$ - $\delta_{\mathrm{CP}}$) plane as compared to their individual reach.
}


\arxivnumber{2408.aaaaa}

\begin{document}
\maketitle

\section{Introduction and motivation}
\label{sec:introducion}

\begin{figure}[t!]
	\centering
	\includegraphics[width=\linewidth]{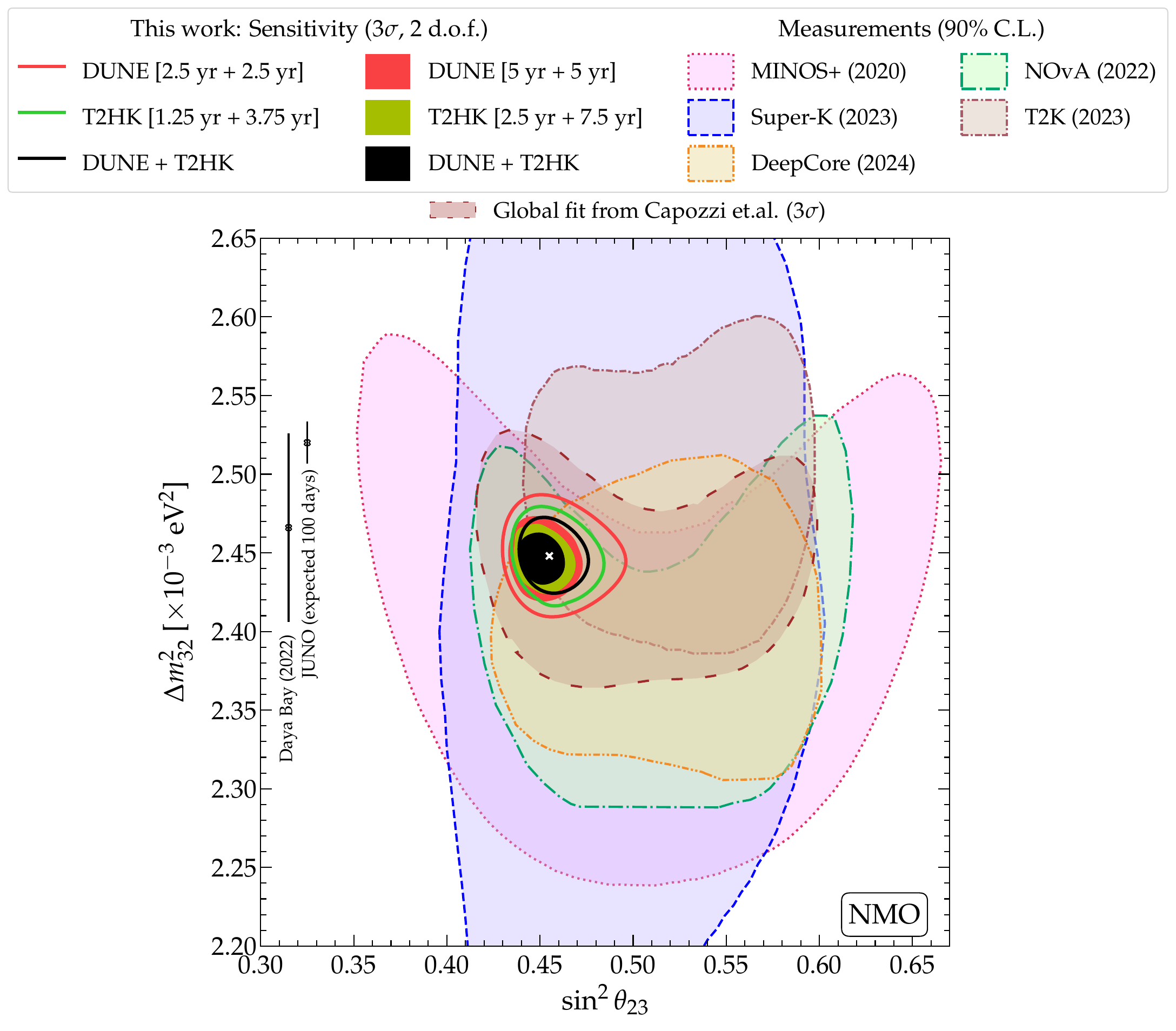}
	\caption{\footnotesize{Allowed ranges at 3$\sigma$ (2 d.o.f.) in the atmospheric mixing parameters, $\sin^{2}\theta_{23}$ and $\Delta m^{2}_{32}$, using DUNE, T2HK, and DUNE + T2HK. DUNE is expected to have an exposure of 480 kt$\cdot$MW$\cdot$yr and T2HK, an exposure of 2431 kt$\cdot$MW$\cdot$yr. We assume DUNE (T2HK) running for 5 (2.5) yr in $\nu$ and 5 (7.5) yr in $\bar{\nu}$ mode while estimating nominal exposure. We also depict the allowed ranges for the same when only half of the total projected exposure is considered (DUNE: [2.5 yr in $\nu$ + 2.5 yr in $\bar{\nu}$], T2HK: [1.25 yr in $\nu$ + 3.75 yr in $\bar{\nu}$]). Existing allowed ranges are from: Super-Kamiokande~\cite{Super-Kamiokande:2023ahc}, Tokai-to-Kamioka (T2K)~\cite{T2K:2023smv}, NuMI Off-Axis $\nu_{e}$ Appearance (NO$\nu$A)~\cite{NOvA:2021nfi}, Main injector neutrino oscillation search (MINOS+)~\cite{MINOS:2020llm}, and IceCube DeepCore~\cite{IceCube:2024xjj,Yu:2023tmw} at 90\% C.L. We also show existing (expected) bounds on $\Delta m^{2}_{32}$ from Daya Bay~\cite{DayaBay:2022orm} 
 (Jiangmen Underground Neutrino Observatory; JUNO ~\cite{JUNO:2022mxj}).  To have a complete picture, we also show the present global fit, following Ref.~\cite{Capozzi:2021fjo} at 3$\sigma$ (1dof). \textit{Our projected allowed ranges improve from the existing bounds by manifold.} See appendix~\ref{appendix1} for a detailed discussion. Further, figure~\ref{fig:allowed-ranges-nmo} elaborates on the importance of both neutrino and antineutrino modes in this sensitivity separately. Also, see figure~\ref{fig:appendix-allowed-ranges-nmo} for a few additional sensitivity curves. }
 } 
 \label{fig:moneyplot}
 \end{figure}

Current oscillation experiments in the three-neutrino paradigm depict potent complementarity. The Solar experiment's precision measurements of solar mixing angle have been combined with KamLAND's ability to determine solar mass splitting well, enabling their synergy to be utilized in the neutrino community for a very long time. The combination of these data and those from long-baseline (LBL) accelerator and atmospheric neutrinos represents the minimal dataset for the characterization of oscillation parameters. The unprecedented precision obtained on the reactor mixing angle ($\theta_{13}$) from short-baseline reactor data like Daya Bay~\cite{DayaBay:2022orm} has further alleviated the uncertainty in the measurements of other unknowns like $\theta_{23},\, \delta_{\mathrm{CP}}$, and the sign of $\Delta m^{2}_{31}$ indirectly by reducing the correlations. The atmospheric neutrino data, when complemented with the long-baseline data, allows us to gain valuable insight into the atmospheric parameters. For example, Super-K, with its extensive statistics, can impose strict constraints on the measurements of $\theta_{23}$, while MINOS/MINOS$+$, benefiting from a precisely known $L/E$ ratio, can offer more accurate measurements of the atmospheric mass splitting, as illustrated in figure~\ref{fig:moneyplot}. In this context, recently, the Super-K experiment has enhanced its precision on atmospheric neutrino oscillation parameters by using the number of tagged neutrons to enhance the separation between neutrino and antineutrino events, improving the efficiency to classify the multi-ring events using a boosted decision tree algorithm, and adding 48\% exposure by analyzing events from an expanded fiducial volume and from 1186 additional live-days, including the data which were collected after a major refurbishment of the detector in 2018~\cite{Super-Kamiokande:2023ahc}.

The presently running accelerators~\cite{Ali:2022mrp,Catano-Mur:2022kyq} and atmospheric experiments~\cite{sk,IceCube:2019dyb} along with the high-precision measurements from reactors~\cite{DayaBay:2018yms,Seo:2019shs} have helped to achieve the current relative 1$\sigma$ precision of about 1.1\% and 6.7\% in the atmospheric parameters: $\Delta m^{2}_{31}$ and $\sin^2\theta_{23}$, respectively~\cite{Capozzi:2021fjo}. The upcoming medium-baseline reactor oscillation experiment JUNO~\cite{JUNO:2022mxj} is expected to achieve considerable improvement in the precision of atmospheric mass-squared difference $\Delta m^2_{32}$ as compared to the current precision from Daya Bay~\cite{DayaBay:2022orm} reactor experiment, as can be seen from figure~\ref{fig:moneyplot}. The DeepCore array consisting of 8 dedicated strings with denser spacing in the central region of IceCube has enabled the detection and reconstruction of atmospheric neutrinos with energies as low as a few GeV, providing high-precision measurements of 2-3 oscillation parameters. Using convolutional neural networks with 9.3 years of data, the IceCube DeepCore has provided a new high-precision measurements of $\Delta m^{2}_{32} = 2.40^{+0.05}_{-0.04} \times 10^{-3}$  eV$^{2}$ and $\sin^{2}\theta_{23}= 0.54^{+0.04}_{-0.03}$~\cite{IceCube:2024xjj} assuming normal mass ordering (NMO), which are compatible and complementary with the existing measurements from the long-baseline experiments. A new extension of IceCube, namely the IceCube Upgrade to be deployed in the polar session of 2025/26 with seven new strings in the central region of DeepCore detector and an energy threshold of around 1 GeV is expected to improve the precision of 2-3 oscillation parameters by (20-30)\%~\cite{IceCube:2023ins}. In ref.~\cite{IceCube-Gen2:2019fet}, the combined sensitivity of the future JUNO, IceCube Upgrade, and PINGU data was estimated to resolve the pressing issue of neutrino mass ordering. The under-construction water Cherenkov neutrino detector KM3NeT/ORCA also has the potential to shed light on neutrino mass ordering and 2-3 oscillation parameters~\cite{KM3NeT:2021ozk}. In fact, recently, they announced their measurements of 2-3 oscillation parameters using an initial configuration with 6-detection units of photo-sensors corresponding to an exposure of 433 kt$\cdot$yr, collected in 510 days of data taking~\cite{KM3NeT:2024ecf}. They reveal a best-fit of $\sin^{2}\theta_{23}= 0.51$ and $\Delta m^{2}_{31} = 2.14 \times 10^{-3}$  eV$^{2}$ with their initial configuration, referred to as ORCA6. Further, in the Neutrino 2024 conference, the KM3NeT/ORCA collaboration showed slightly improved measurements of 2-3 oscillation parameters using an updated exposure of 715 kt$\cdot$yr~\cite{coelho_orca}. In ref.~\cite{KM3NeT:2021rkn}, a combined analysis of the prospective JUNO and KM3NeT/ORCA data was performed to determine the correct neutrino mass ordering.

The present hints of non-maximal $\theta_{23}$ from the global oscillation data~\cite{Capozzi:2021fjo,deSalas:2020pgw,Esteban:2020cvm,NuFIT} give rise to two probable solutions: $\theta_{23} < 45^{\circ}$ or the lower octant solutions (LO) and $\theta_{23} > 45^{\circ}$ or the higher octant solutions (HO)~\cite{Fogli:1996pv, Barger:2001yr, Minakata:2002qi, Minakata:2004pg,Hiraide:2006vh}. Prior to delving into the determination of the correct octant, it is imperative to eliminate the possibility of maximal mixing with a high level of confidence. This investigation was undertaken earlier, taking into account the forthcoming DUNE (Deep Underground Neutrino Experiment)~\cite{Agarwalla:2021bzs}. In this study, we explore the synergies between the two prominent upcoming neutrino experiments: DUNE~\cite{DUNE:2020jqi,DUNE:2020ypp}, designed to receive a wide-band on-axis neutrino beam traversing a distance of around 1300 km with substantial Earth matter effect, and T2HK (Tokai to Hyper-Kamiokande)~\cite{Hyper-KamiokandeProto-:2015xww,Hyper-Kamiokande:2018ofw}, which proposes the use of a narrow-band off-axis neutrino beam with a baseline of 295 km having minimal Earth matter effect. We investigate how the collaborative synergy of these setups enhances their individual sensitivities. Additionally, we assess the expected sensitivities of currently running long-baseline experiments: T2K (Tokai to Kamioka)~\cite{T2K:2011qtm} and NO$\nu$A (NuMI Off-Axis $\nu_{e}$ Appearance)~\cite{NOvA:2007rmc}, considering their full projected exposures. Our findings indicate that the combined DUNE + T2HK configuration can detect a departure from maximal $\theta_{23}$ with exceptional significance. Moreover, their complementarity turns out to be essential for achieving degeneracy-free and significantly precise measurements of both $\sin^2\theta_{23}$ and $\Delta m^{2}_{31}$.

As mentioned before, figure~\ref{fig:moneyplot} shows a part of our main results exhibiting the expected precision in 2-3 oscillation parameters: $\sin^{2}\theta_{23}$ and $\Delta m^{2}_{32}$ at 3$\sigma$ confidence level (C.L.) using the projected full and half exposures of standalone DUNE and T2HK, and their combination (DUNE + T2HK). To have a better comparison, we also include the currently allowed regions from the ongoing (T2K, NO$\nu$A, IceCube DeepCore, Super-Kamiokande), completed (Daya Bay and MINOS+), and upcoming (JUNO) experiments at 90\% C.L. assuming NMO. The details of these experiments (such as exposures, runtime, and current status) and the best-fit values of the oscillation parameters obtained from them are given in table~\ref{tab:other-experiments}. In figure~\ref{fig:moneyplot}, we also show the allowed region from the global fit of all the available oscillation data following Ref.~\cite{Capozzi:2021fjo} at 3$\sigma$, assuming NMO. More details related to this figure can be found in Appendix~\ref{appendix2}. Note that in figure~\ref{fig:moneyplot}, we compare the sensitivities of various experiments in terms of $\Delta m^{2}_{32}$ (see the y-axis). However, we present all our results in terms of $\Delta m^{2}_{31}$ in the present manuscript.

\begin{table}[htb!]
 \begin{tabular}{ | c | c | c | c | c | c|}
 \hline
 \multirow{3}{*}{Experiment}
 & \multicolumn{2}{c|}{Best-fit} &  \multirow{3}{*}{Exposure} &  \multirow{3}{*}{Span} &  \multirow{3}{*}{Status}\\ 
 \cline{2-3}
 &  \multirow{2}{*}{$\sin^{2}\theta_{23}$} & $\Delta m^{2}_{32}$ &    &  &\\
  & & $[10^{-3}] ~\text{eV}^2$ &   & &\\
  \hline
\multirow{3}{*}{T2K~\cite{T2K:2023smv}} & \multirow{3}{*}{0.56} & \multirow{3}{*}{2.49} &  P.O.T. & \multirow{3}{*}{2010 - 2020} & \multirow{3}{*}{Ongoing}\\
& & & $16.3 \times 10^{20}$ ($\nu$)  & & \\
& & & $19.7 \times 10^{20}$ ($\bar{\nu}$)  & & \\
\hline
\multirow{3}{*}{NO$\nu$A~\cite{NOvA:2021nfi}} & \multirow{3}{*}{0.57} & \multirow{3}{*}{2.41} & P.O.T. &  & \multirow{3}{*}{Ongoing} \\
& & & $13.6 \times 10^{20}$ ($\nu$) & 2016 - 2019 & \\
& & & $12.5 \times 10^{20}$ ($\bar{\nu}$) & 2014 - 2020 & \\
\hline
MINOS/ & \multirow{2}{*}{0.43} & \multirow{2}{*}{2.4} & $23.76 \times 10^{20}$ P.O.T. & 2005 - 2016 & \multirow{2}{*}{Completed}\\
MINOS$+$~\cite{MINOS:2020llm}& & & 60.75 kt$\cdot$yr & 2011 - 2016 & \\
\hline
Super-K & \multirow{2}{*}{0.49} & \multirow{2}{*}{2.4} & \multirow{2}{*}{484.2 kt$\cdot$yr}  & \multirow{2}{*}{1996 - 2020} & \multirow{2}{*}{Ongoing}\\ 
(I - V)~\cite{Super-Kamiokande:2023ahc} & & & & & \\
\hline
IceCube- & \multirow{2}{*}{0.54} & \multirow{2}{*}{2.4} & \multirow{2}{*}{9.3 yr} & \multirow{2}{*}{2012 - 2021} & \multirow{2}{*}{Ongoing}\\
DeepCore~\cite{IceCube:2024xjj} & & & & & \\
\hline
Daya Bay~\cite{DayaBay:2022orm} & - & 2.466 & 3158 live days & 2011 - 2020 & Completed\\
\hline
JUNO~\cite{JUNO:2022mxj} & - & 2.52 & 6 yr & - & Upcoming\\
\hline
 \end{tabular}
 \caption{Existing and expected best-fit values, collected exposure, runtime, and the present status of atmospheric parameters using long-baseline, atmospheric, and reactor experiments. Collected exposure is expressed in protons on target (P.O.T.) for long-baseline experiments and in kt$\cdot$yr for atmospheric experiments. For certain cases, we give the number of years of data collection. Using these details, currently allowed regions in the $\sin^{2}\theta_{23}$ and $\Delta m^{2}_{32}$ plane is shown in figure~\ref{fig:moneyplot}. 
 }
 \label{tab:other-experiments}
\end{table}

The other main results include sensitivity towards deviation from maximal $\sin^{2}\theta_{23}$, exclusion of wrong octant solutions of $\sin^{2}\theta_{23}$, and precision on atmospheric parameters, studied as a function of exposure. We find that the \textit{complementarity between DUNE and T2HK plays a crucial role in reducing the dependency on large projected exposures of standalone experiments by manifold. Furthermore, the synergy between them helps in removing the degeneracies introduced by the individual setups, if any. } 

The manuscript is laid out as follows. In Sec. \ref{sec:events}, we summarize the characteristic features of DUNE and T2HK and discuss the effect of wrong-sign contaminations and variation of 2-3 oscillation parameters on the total event statistics and event spectra, respectively. Next, Sec. \ref{sec:results}, elaborates on our results and discussions. We compute the sensitivities in establishing deviation from maximal $\sin^2\theta_{23}$, exclusion of wrong octant solutions of $\sin^2\theta_{23}$, and the precision of $\sin^2\theta_{23}$ and $\Delta m^2_{31}$, using DUNE and T2HK in both isolation and combination. We also analyze the effect of scaled exposure on the above-mentioned sensitivity studies. Then, Sec. \ref{sec:contour} shows projected allowed ranges in $(\sin^2\theta_{23}-\delta_{\mathrm{CP}})$ plane, using half and full exposures in DUNE, T2HK, and their combination. Finally, in Sec. \ref{conclusion}, we summarize our findings and provide concluding remarks. Additionally, we have two appendices. While Appendix \ref{appendix1} comprehends the individual roles of $\nu$ and $\bar{\nu}$ modes, using  DUNE, T2HK, and DUNE + T2HK in $(\sin^2\theta_{23}-\delta_{\mathrm{CP}})$ plane; Appendix \ref{appendix2} depicts past, present, and upcoming projected sensitivities in $(\sin^2\theta_{23}-\delta_{\mathrm{CP}})$ plane.

\section{Experimental details and total event rates}
\label{sec:events}

We initiate our discussion by comparing and contrasting the two upcoming long-baseline experiments under consideration: DUNE and T2HK. Following this, we compute the expected total appearance and disappearance event rates in both the $\nu$ and the $\bar{\nu}$ modes for the presently allowed 3$\sigma$ ranges in $\theta_{23}$ and $\Delta m^{2}_{31}$~\cite{Capozzi:2021fjo} using GLoBES~\cite{Huber:2004ka, Huber:2007ji}. Since the far detectors in both DUNE and T2HK are unable to differentiate between neutrinos and antineutrinos, we also discuss the effect of ``wrong-sign'' contamination, which is considered a part of the signal in both the experiments.
%
\subsection{Complementarities between DUNE and T2HK}
\label{sec:2a}
 
DUNE  and T2HK  are two promising long-baseline experiments expected to achieve significant aspects of physics beyond the three-neutrino oscillations. We consider a single-phase state-of-the-art 40 kt Liquid Argon Time Projection Chamber (LArTPC) far detector in DUNE and a 187 kt Water Cherenkov far detector in T2HK as referred in their cumulative design reports, respectively~\cite{DUNE:2021tad,Hyper-Kamiokande:2016srs}. The neutrino flux in DUNE is expected to be wide-band on-axis, ranging from a few hundreds of MeV to a few tens of GeV, peaking at 2.5 GeV. This wide-band nature enables DUNE access to an envelope of various $L/E$ ratios, where $L$ corresponds to the distance that neutrino travels from source to detector and $E$ refers to the neutrino beam energy. Contrastingly, T2HK is expected to use a 2.5$^{\circ}$ off-axis J-PARC neutrino beam, the flux expected to peak at 0.6 GeV. The higher baseline in DUNE (1285 km; from Fermilab to South Dakota) ensures sufficient matter effect, while the relatively shorter baseline in T2HK (295 km; from J-PARC proton synchrotron facility to Hyper-Kamiokande) secures better precision in measurements of the intrinsic CP phase. The line-averaged constant Earth matter density ($\rho_{\mathrm{avg}}$) in DUNE is considered to be 2.848 g/cm$^3$, while in T2HK, it is taken as 2.8 g/cm$^3$. In DUNE, we consider 2\% detector systematic uncertainties in the appearance and  5\% in disappearance signal events following ref.~\cite{DUNE:2021cuw}. The binned events in T2HK have been matched with ref.~\cite{Hyper-Kamiokande:2016srs}, considering 5\% in appearance and 3.5\% in disappearance systematic uncertainties in signal events. Recently, apart from considering 5\% of conservative systematic uncertainties in appearance events, the T2HK collaboration is expecting to improve this uncertainty to about 2.7\% by the time they start taking their data in real-time~\cite{Munteanu:2022zla}. Thus we compare our results with both of these choices in figure~\ref{fig:octant exclusion}. For the runtime in DUNE, they expect to witness a balanced run between neutrino and antineutrino modes following [5 years in $\nu$ + 5 years in $\bar{\nu}$], T2HK aims instead to have an almost equal number of events in both the modes, thus following the 1:3 ratio of [2.5 years in $\nu$  + 7.5 years in $\bar{\nu}$]. Owing to the much higher detector fiducial mass, T2HK expects to accumulate about $2.7 \times 10^{22}$ P.O.T. per yr, providing a benchmark exposure of 2431 kt$\cdot$MW$\cdot$yr, while DUNE envisions a P.O.T. of around $1.1 \times 10^{21}$ per year with a benchmark exposure of 480 kt$\cdot$MW$\cdot$yr. In table~\ref{table:one}, we mention our assumed benchmark values and the ranges, which are taken from ref.~\cite{Capozzi:2021fjo}.

As the far detector deployment schedule and beam power scenarios are both subject to change, the results shown in this work are consistently given in terms of exposure in units of kt$\cdot$MW$\cdot$yr, which is agnostic to the exact staging scenario but can easily be expressed in terms of experiment years for any desired scenario. For having a complete summary, we present the sensitivity studies of DUNE and T2HK along with the full potential of ongoing long-baseline experiments: T2K and NO$\nu$A. We present our findings using the entire projected exposures of $84.4$ kt$\cdot$MW$\cdot$yr, generating $7.8 \times 10^{21}$ P.O.T. with a 750 kW beam power, evenly distributed between neutrino and antineutrino modes, as outlined in the ongoing long-baseline experiment T2K~\cite{T2K:2014xyt}. Additionally, we conduct simulations for the full projected exposure of NO$\nu$A, amounting to 58.8 kt$\cdot$MW$\cdot$yr and producing $3.6 \times 10^{21}$ P.O.T. with a 700 kW beam power, equally divided between neutrino and antineutrino modes, in accordance with ref.~\cite{NOvA:2007rmc,Patterson:2012zs}. In both experiments, we assume uncorrelated 5\% and 10\% systematic errors on signal and background events for both appearance and disappearance event rates.
%
\begin{table}[t!]
\centering
\resizebox{\columnwidth}{!}{%
\begin{tabular}{|c|c|c|c|c|c|c|}
\hline 
\multirow{2}{*}{\textbf{Parameter}} & $\Delta m^2_{21}/10^{-5}$ & \multirow{2}{*}{$\sin^{2}\theta_{12}/10^{-1}$} & \multirow{2}{*}{$\sin^{2}\theta_{13}/10^{-2}$} & \multirow{2}{*}{$\sin^2\theta_{23}/10^{-1}$} & $\Delta m^2_{31}/10^{-3}$  & $\delta_{\text{CP}}$ \\
				&($\mathrm{eV^{2}}$) & & & &($\mathrm{eV^{2}}$)&($^\circ$)\\
				\hline 
				{\textbf{Benchmark}} & $7.36$ & $3.03$ & $2.23$ & $4.55$ & $2.522$ & $223$\\
    \hline
				\textbf{$3\sigma$ range}& - & - & - & 4.16 - 5.99 & 2.436 - 2.605 & 139 - 355\\
					\hline 
			\end{tabular}
			}
			\caption{\footnotesize{The benchmark values of the oscillation parameters and their corresponding 3$\sigma$ allowed ranges considered in our study assuming normal mass ordering (NMO) following the ref.~\cite{Capozzi:2021fjo}.}}
			\label{table:one}
	\end{table}

\subsection{Events due to wrong-sign contamination }
\label{sec:2c}
\begin{table}[t!]
 \centering
 \begin{tabular}{ | c  *{6}{>{\centering\arraybackslash}p{2.2cm} |}}
 \hline
 \multicolumn{2}{|c|}{\multirow{3}{*}{Experiment}}
 & \multicolumn{4}{c|}{Number of events (NMO)} \\
 \cline{3-6}
          & & \multicolumn{2}{c|}{Appearance} 
          & \multicolumn{2}{c|}{Disappearance} \\
          \cline{3-6}
          & & $\nu$ mode & $\bar{\nu}$ mode & $\nu$ mode & $\bar{\nu}$ mode \\
 \hline
  \multirow{4}{*}{DUNE} & w/  & \multirow{2}{*}{1592} & \multirow{2}{*}{294} & \multirow{2}{*}{14598} & \multirow{2}{*}{8270} \\
                         & wrong-sign    &  &  &  & \\            
                      & w/o    & \multirow{2}{*}{1576} & \multirow{2}{*}{186} & \multirow{2}{*}{13413} & \multirow{2}{*}{4360}   \\
                 & wrong-sign &  &  & &  \\
 \hline
 \multirow{4}{*}{T2HK} & w/  & \multirow{2}{*}{1598} & \multirow{2}{*}{919} & \multirow{2}{*}{10064} & \multirow{2}{*}{13949} \\
                 &  wrong-sign   & &  & & \\
                       & w/o    & \multirow{2}{*}{1588}  & \multirow{2}{*}{755} & \multirow{2}{*}{9487} & \multirow{2}{*}{8985}   \\
                       &  wrong-sign    &  & & &  \\
 \hline
 \end{tabular}
 \caption{Total (Signal) appearance and disappearance event rates in DUNE and T2HK assuming 480 kt$\cdot$MW$\cdot$yr and 2431 kt$\cdot$MW$\cdot$yr of exposure, respectively.  We fix the values of the standard mixing parameters to their benchmark values from~\cite{Capozzi:2021fjo}; see table~\ref{table:one}. }
 \label{table:two}
\end{table}
%

\begin{figure}[ht!]
	\centering
	\includegraphics[width=\linewidth]{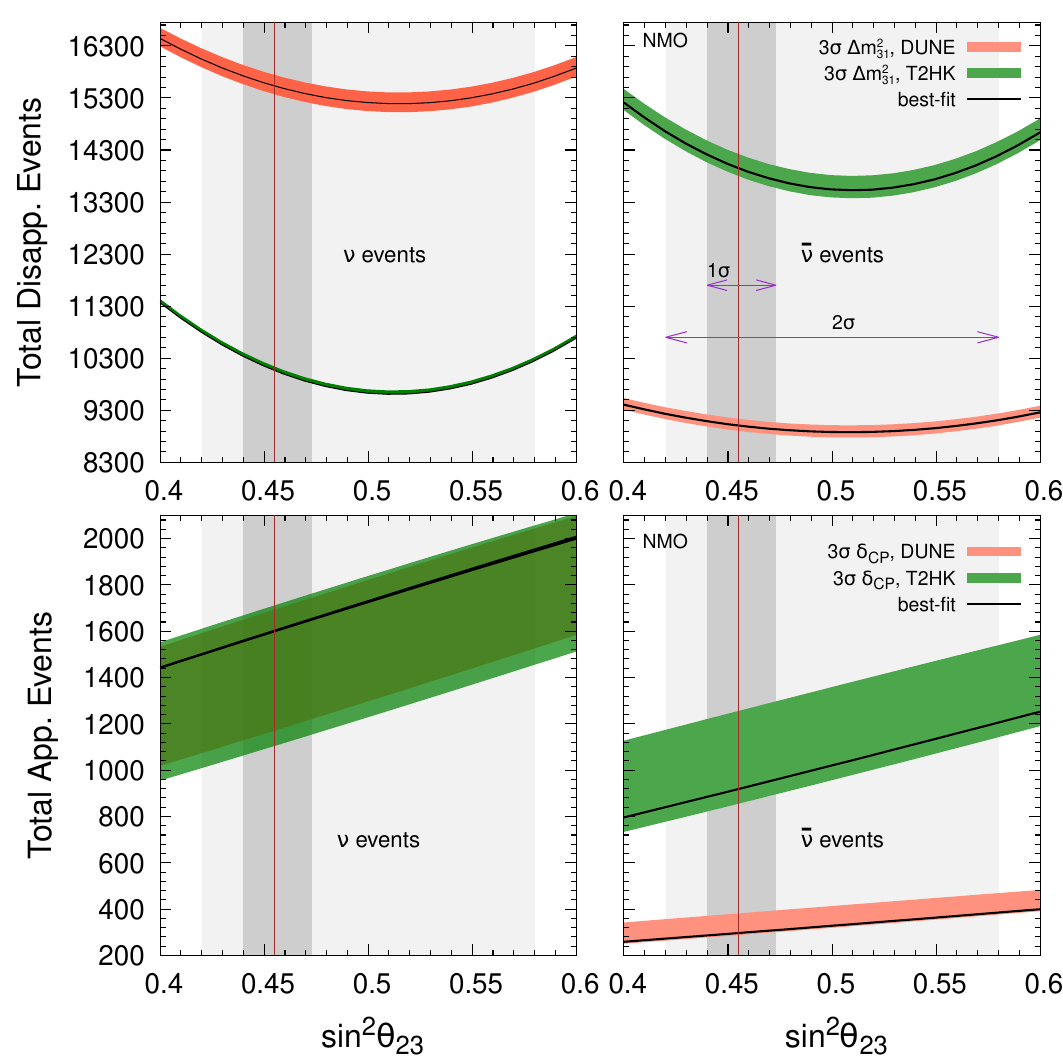}
	\caption{\footnotesize{Total (only signal) disappearance and appearance event rates as a function of $\sin^2\theta_{23}$ in DUNE and T2HK assuming NMO. The left and right panel depicts the same in neutrino and antineutrino modes for 480 kt$\cdot$MW$\cdot$yr and 2431 kt$\cdot$MW$\cdot$yr of exposure in DUNE and T2HK, respectively. Band depicts $~3\sigma$ uncertainty in $\Delta m^2_{31}$ (upper panel) and $\delta_{\rm CP}$ (lower panel). 
 }}
	\label{fig:1}
\end{figure}
In principle, reconstructing the charge of muons (to identify and segregate neutrinos from antineutrinos), event-by-event, is not feasible in DUNE and T2HK. There is always the occurrence of ``wrong-sign'' $ \bar{\nu}_{\mu}\, (\nu_{\mu})$ charged-current (CC) events when the primary beam is $\nu_{\mu} (\bar{\nu}_{\mu})$. Similarly, there is the contamination of ``wrong-sign'' $ \bar{\nu}_{e}\, (\nu_{e})$ charged-current (CC) events in neutrino (antineutrino) modes.

For the energy range under consideration, there is suppression in both the antineutrino cross section and flux than a neutrino, which leads to a relatively more contribution of wrong-sign $\nu_{\mu}$ CC events in antineutrino mode than its counterpart. Apart from this, in Nature, positive mesons are more abundant than their negative counterpart as they are produced following the $pp$ or the $pn$ collisions~\cite{Dore:2018ldz}. Therefore, the neutrino beam is more intense than the antineutrino beam, and hence, the contamination of wrong-sign neutrinos in the antineutrino beam is higher. Both DUNE and T2HK follow horn current terminology, where the neutrino-enhanced beam is coined as forward horn current (FHC), and the antineutrino-enriched beam is the reverse horn current (RHC). In FHC, the wrong-sign flux is concentrated in the high-energy tail of the flux spectrum, where leptons are more likely to be forward and energetic due to the kinematics of neutrino and antineutrino scattering, while in RHC, they are concentrated around the low reconstructed energies~\cite{Katori:2016yel,DUNE:2020jqi}. Therefore, in general, we expect that the number of $\bar{\nu}$ events in $\nu$ beam to be way lesser than contamination of ${\nu}$ beam with $\bar{\nu}$.

In table \ref{table:two}, we compute the total event rates (signal) in $\nu_\mu\rightarrow\nu_\mu$ disappearance channel and $\nu_\mu\rightarrow\nu_{e}$ appearance channel for both neutrino and antineutrino modes in DUNE and T2HK, using benchmark oscillation parameters (refer to table~\ref{table:one}) with and without the inclusion of wrong-sign events. From the illustrative events shown in table~\ref{table:two} under two scenarios, we observe that the contribution from wrong-sign events is more in the RHC ($\bar{\nu}$ mode) than in FHC ($\nu$ mode). Moreover, in RHC, this contamination is more for the disappearance event rates (~50\% in DUNE, 35\% in T2HK) than appearance rates (~36\% in DUNE and ~17\% in T2HK). Consequently, given its larger size, T2HK is likely to exhibit greater effects from this contamination than DUNE. Following the general convention, in all our analyses henceforth, we have considered the wrong-sign contributions in both FHC and RHC signal events for both DUNE and T2HK.

\subsection{Total appearance and disappearance event rates}
\label{sec:2d}

Figure~\ref{fig:1} illustrates the total disappearance and appearance event rates in DUNE and T2HK. While the disappearance rates are affected mostly by uncertainty in $\Delta m^{2}_{31}$ and $\sin^{2}\theta_{23}$, the uncertainty in $\delta_{\rm CP}$ and $\sin^{2}\theta_{23}$ affects appearance rates predominantly, therefore we show bands of currently allowed 3$\sigma$ in $\Delta m^2_{31}$ for disappearance and $\delta_{\rm CP}$ in appearance event rates. The disappearance rates follow a U-shaped distribution when studied as a function of $\sin^{2}\theta_{23}$. This is because the disappearance rates $\propto (1-\sin^{2}\theta_{23})$. This points towards multiple combinations of $\sin^{2}\theta_{23}-\Delta m^2_{31}$ with same number of events~\cite{Agarwalla:2021bzs,Agarwalla:2022xdo}. Further, we observe that for values of $\sin^{2}\theta_{23}$ in the HO but close to 0.5, the curves show a flat behavior in both DUNE and T2HK. This hints that for these values of $\sin^{2}\theta_{23}$, sensitivity towards deviation from maximality will mostly come from the appearance rates, while disappearance rates may dominate later. Also, the minimum is not exactly at 0.5; instead, it is seen slightly shifted towards HO due to finite $\theta_{13}$ correction~\cite{Raut:2012dm}. Following higher runtime, more expected flux in neutrino mode, and substantial matter effect, we expect higher neutrino statistics in DUNE than T2HK, keeping NMO fixed. Similarly, higher runtime in antineutrino mode for T2HK implies higher antineutrino statistics. Having access to a wide-band beam makes DUNE capable of analyzing several $L/E$ ratios and more susceptible to a change in the value of $\Delta m^{2}_{31}$, unlike T2HK. This explains the higher neutrino disappearance statistics and a wider band when we vary $\Delta m^{2}_{31}$ in DUNE relative to T2HK. Further, DUNE's access to both the first and second oscillation maxima assures high disappearance rates in neutrino mode~\cite{DUNE:2020jqi} than T2HK, which has relatively fewer events at the second oscillation maximum~\cite{Hyper-Kamiokande:2018ofw}. In antineutrino mode, a higher runtime overcomes cross section suppression in T2HK, which is not observed in the case of DUNE. Considering the appearance events in neutrino mode, DUNE has a higher runtime but lesser exposure, while T2HK has a lesser runtime but higher exposure; therefore, both experiments have almost similar event rates. In contrast, T2HK has higher statistics in $\bar{\nu}$ mode due to more runtime than DUNE.

Below, we show how the above contrasting features grant DUNE and T2HK the capability to probe atmospheric parameters organically, complementing each other.

\section{Projected sensitivities and its variation with total exposure}
\label{sec:results}
We project the expected sensitivities in DUNE, T2HK, and their combination to study the atmospheric parameters based on the detailed computation of event rates discussed above. Our results and analyses are based upon the current scenario in 3$\nu$ paradigm from the global oscillation data answering three crucial questions: (i) establishing deviation from maximal $\theta_{23}$, (ii) precision measurements in the 2-3 sector; $\sin^2\theta_{23}$ and $\Delta m^2_{31}$\,, and (iii) rejecting the wrong octant solutions of $\sin^2\theta_{23}$. Following the definition of Poissonian $\chi^2$~\cite{Baker:1983tu}, we estimate the median sensitivity~\cite{Cowan:2010js} of a given experiment in the frequentist approach~\cite{Blennow:2013oma} as  
\begin{equation}
\chi^2 (\vec{\omega}, \, \kappa_{s}, \, \kappa_{b,l})= \underset{( \vec{\lambda}, \, \kappa_{s}, \,\kappa_{b,l})}{\mathrm{min}}\left\{  2\sum_{i=1}^{n}(\tilde{y_i}-x_i-x_i\mathrm{ln}\frac{\tilde{y_i}}{x_i})+\kappa^2_{s}+ \sum_{l}\kappa^2_{b,l}\right\}\, , 
\label{eq:chi2} 
\end{equation}
where $n$ is the total number of reconstructed energy bins and $\vec{\lambda}$ is the set of oscillation parameters that are marginalized in the fit. The choice of set $\vec{\lambda}$ is discussed later in every subsection. Further,
\begin{equation}
\tilde{y_i}\,(\vec{\omega},\{\kappa_{s},\kappa_{b,l}\}) = N^{th}_i(\vec{\omega})[1+\pi^s\kappa_{s}] + \sum_{l}N^b_{i,l}(\vec{\omega})[1+\pi^b\kappa_{b,l}]\, .
\label{chi}
\end{equation}
Here, $N^{th}_i\,(\vec{\omega})$ is the number of signal events predicted in the $i$-th energy bin for a given set of oscillation parameters $\vec{\omega} = \left\lbrace \theta_{23}\,,\theta_{13}\,,\theta_{12}\,,\Delta \mathrm{m}^{2}_{21}\,, \Delta \mathrm{m}^{2}_{31}\,,\delta_{\mathrm{CP}}\right\rbrace$. $N^{b}_{i,l}\,(\vec{\omega})$ denotes the number of background events in the $i$-th energy bin where the neutral current (NC) backgrounds are independent of the oscillation parameter $\vec{\omega}$, while the charged current (CC) backgrounds is dependent on the oscillation parameters. $\pi^s$ is the pull term for systematic uncertainty on signal events. $\pi_{b,l}$ is the pull term for the systematic uncertainties on the $l$-th background contribution for any given channel in signal. These pull terms are uncorrelated with one another and have the same values in neutrino and antineutrino modes.  We incorporate the corresponding data in Eq.~\ref{eq:chi2} using the variable $x_{i}= N^{ex}_i\, +\, N^b_{i,l}$, where $N^{ex}_i$ indicates the observed CC signal events in the $i$-th energy bin via $\nu_{\mu}\rightarrow \nu_{\mu}$ and $\bar{\nu}_{\mu}\rightarrow \bar{\nu}_{\mu}$ disappearance channels and $\nu_{\mu}\rightarrow \nu_{e}$ and $\bar{\nu}_{\mu}\rightarrow \bar{\nu}_{e}$ appearance channels. Here, $N^b_{i,l}$  represents the $l$-th background contribution for a given channel. Throughout the simulation, we use publicly available software GLoBES~\cite{Huber:2002mx,Fogli:2002pt,Gonzalez-Garcia:2004pka}. We fix the mass ordering to NMO while generating data, as there are weak hints from global oscillation data favoring NMO at $\sim 2.5\sigma$~\cite{Capozzi:2021fjo,NuFIT}. In the fit, we marginalize over the allowed regions in oscillation parameters as mentioned in table~\ref{table:one}. We do not include any correlations among them as by the time these future experiments start taking data, these correlations will likely weaken~\cite{Song:2020nfh}. We consider the benchmark choices for $\theta_{12}$ and $\theta_{13}$ fixed~\cite{Capozzi:2021fjo}, as we do not expect the precision (2.8\%) achieved by Daya Bay to improve in the coming years~\cite{DayaBay:2022orm}. Although the present-day uncertainty in $\theta_{12}$ is 4.5\% ~\cite{Capozzi:2021fjo}, we do not expect the sensitivity in our study to get affected by it. We also fix the mass ordering in the fit as in the next decade, DUNE is expected to determine the mass ordering within initial years of data.

\subsection{Establishing deviation from maximal $\sin^{2}\theta_{23}$}
\begin{figure}[htb!]
	\centering
	\includegraphics[width=0.8\linewidth]{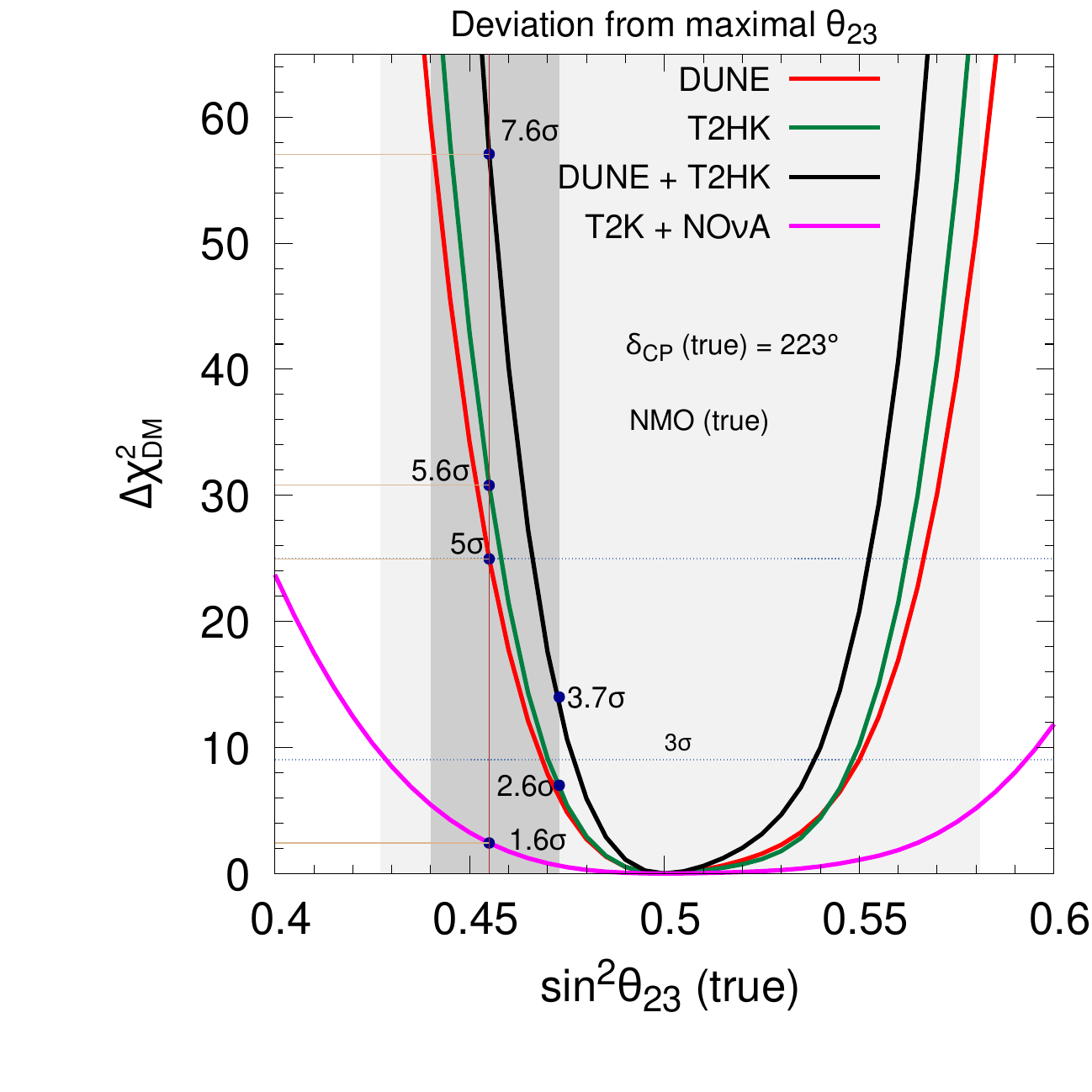}
	\caption{\footnotesize{ Sensitivity of DUNE, T2HK, DUNE + T2HK, and T2K [2.5 yr ($\nu$) + 2.5 yr ($\bar{\nu}$)] + NO$\nu$A [3 yr ($\nu$) + 3 yr ($\bar{\nu}$)] to exclude non-maximal solutions of $\sin^{2}\theta_{23}$ as a function of $\sin^2\theta_{23}$ in the data. For benchmark choices of oscillation parameters and the allowed ranges in $\Delta m^{2}_{31} $ and $\delta_{\rm CP}$ over which we marginalize in the fit, refer to table~\ref{table:one}. We assume exposures of 480 kt$\cdot$MW$\cdot$yr in DUNE, 2431 kt$\cdot$MW$\cdot$yr in T2HK, $84.4$ kt$\cdot$MW$\cdot$yr in T2K, and 58.8 kt$\cdot$MW$\cdot$yr in NO$\nu$A. For illustrative purpose, the benchmark choices of $\sin^{2}\theta_{23}$ is shown by a vertical brown line, projecting out the statistical confidence at the intersection with each curve. \textit{If in Nature, $\sin^{2}\theta_{23}$ is around the lower value of current 1$\sigma$ uncertainty ($\sim 0.473$), then the combination is the only solution to achieve $3\sigma$ with current benchmark values.} 
 }} 
	\label{fig:DM}
\end{figure}
We compute the statistical confidence with which DUNE, T2HK, and DUNE + T2HK can establish a deviation from maximal $\sin^{2}\theta_{23}$ by following 
\begin{equation}
\Delta \chi^2_{\text{DM}}= \underset{\delta_{\mathrm{CP}}\,,\,\Delta m^{2}_{31}}{\mathrm{min}}\left\{ \chi^2\left(\sin^2\theta_{23}^{\mathrm{test}} = 0.5\right)-\chi^2\left(\sin^2\theta_{23}^{\mathrm{true}} \in [0.4,0.6]\right) \right\},
\label{eq:deviation-from-maximality-chi2}
\end{equation}
where, $\vec{\lambda} = \{\delta_{\mathrm{CP}}, \, \Delta m^{2}_{31}\}$ is the set of oscillation parameters over which $\Delta\chi^2_{\rm DM}$ gets marginalized in the fit. So we generate the data for allowed uncertainty in $\sin^{2}\theta_{23}$ (refer to table~\ref{table:one}), while fixing it to 0.5 in the fit. There have been previous studies along this direction in ref.~\cite{Ballett:2016daj}\,, however here we consider the current best-fit values from ref.~\cite{Capozzi:2021fjo} which are similar to other global oscillation studies~\cite{NuFIT,Esteban:2020cvm,deSalas:2020pgw}. We also incorporate the latest collaboration estimates and ancillary files while using GLoBES.
\begin{figure}[t!]
	\centering
	\includegraphics[width=\linewidth]{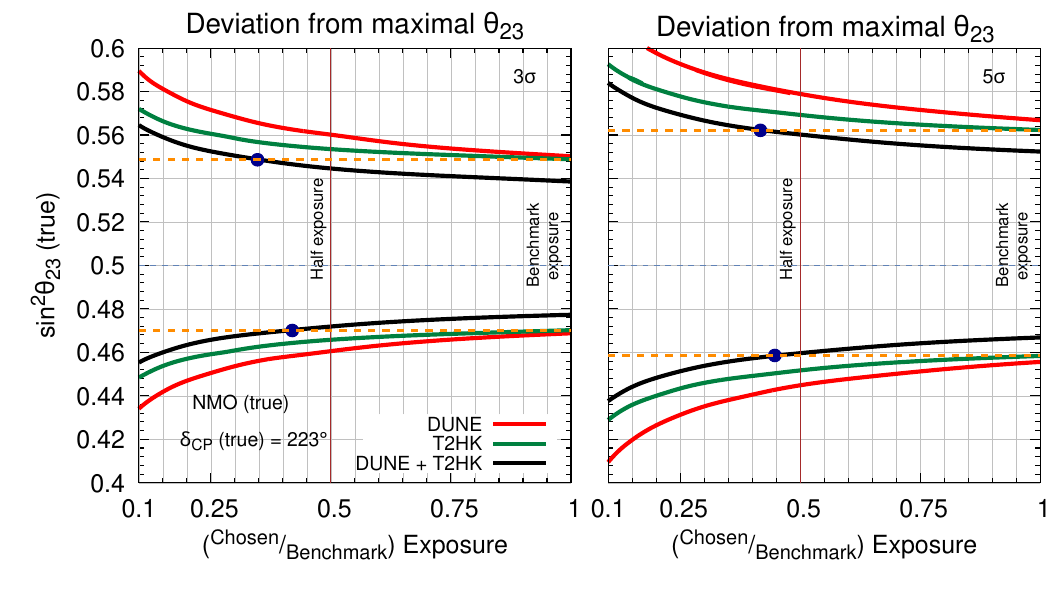}
	\caption{\footnotesize{3$\sigma$  (5$\sigma$) sensitivity of non-maximal $\sin^{2}\theta_{23}$ as a function of scaled (ratio of the chosen to the benchmark value of) exposure, assuming true NMO in the left (right) panel. The ratio reaches 1 at the benchmark choices of the corresponding experimental setups. We generate data by fixing all other oscillation parameters except $\sin^{2}\theta_{23}$ to their best-fit value (see table~\ref{table:one}). We marginalize over $\delta_{\mathrm{CP}}$ and $\Delta m^2_{31}$ in the test-statistics using the ranges of marginalization in table~\ref{table:one}. Dashed orange lines and solid blue circles are used to project and compare the maximum sensitivity attainable by standalone DUNE and T2HK using their nominal exposures with the exposure needed by DUNE + T2HK to achieve the same sensitivity. \textit{We find that using approximately 40\% of their exposures, DUNE + T2HK together achieves comparable sensitivity to each experiment independently running at their nominal exposures.}
  }}
	\label{fig:Exposure_DM}
\end{figure}

Figure~\ref{fig:DM} depicts the sensitivity in establishing deviation from maximality ($\Delta \chi^2_{\rm DM}$ in Eq.~\ref{eq:deviation-from-maximality-chi2}) as a function of true $\sin^{2}\theta_{23}$ for DUNE, T2HK, and their combination. As expected, it is smallest when the true and test equals for $\sin^{2}\theta_{23}$, increasing on both sides as we go away, implying the major contribution from $\sin^{2}2\theta_{23}$ in the leading term of disappearance channel (refer to an elaborate discussion in ref.~\cite{Agarwalla:2021bzs}). However, the U-shape around the $\sin^{2}\theta_{23} = 0.5$ is not symmetric because of the non-zero value of $\theta_{13}$. We observe that even after using the projected full exposure in the present-day long-baseline experiments: T2K and NO$\nu$A; they have lesser sensitivities. One of the major drawbacks in the present-generation experiments is much higher systematic uncertainties in both $\nu_{\mu}\rightarrow \nu_{\mu}$ and $\bar{\nu}_{\mu}\rightarrow \bar{\nu}_{\mu}$ disappearance rates (refer to last paragraph in section~\ref{sec:2a} for corresponding values). The spread of curve around $\sin^{2}\theta_{23} = 0.5$ in all the experimental setups is less when $\sin^{2}\theta_{23} < 0.5$ than $\sin^{2}\theta_{23} > 0.5$. This can be explained due to the finite $\theta_{13}$ corrections~\cite{deGouvea:2005hk,Coloma:2012ut,Raut:2012dm} (Also, refer to discussion around fig.~\ref{fig:1} in Sec.~\ref{sec:2d}). Establishing sensitivity to deviations from maximality primarily depends on the disappearance statistic. In ref.~\cite{Agarwalla:2021bzs}, we also analyze and infer that uncertainty in the values of $\delta_{\mathrm{CP}}$ have minimal impact on this sensitivity. While T2HK, owing to huge disappearance statistics and corresponding lesser expected systematic uncertainties (refer to section~\ref{sec:2a}), is able to achieve better sensitivity than DUNE, DUNE provides better measurements of $\Delta m^{2}_{31}$. Their combination (DUNE + T2HK) makes use of the complementary features among them and achieves a nearly 8$\sigma$ statistical confidence in establishing non-maximal $\sin^{2}\theta_{23}$, considering full exposures in the two experiments and the benchmark values. Furthermore,  \textit{if in Nature, $\sin^{2}\theta_{23}$ is around the upper value of current 1$\sigma$ uncertainty ($\sim 0.473$), then the combination is the only solution to achieve $3\sigma$ with current benchmark values.}

 The product of runtime, fiducial detector mass, and beam power provides the expected experimental exposure. The quantity of exposure is often considered interchangeably with runtime in phenomenological studies of neutrino oscillation. Currently, the DUNE collaboration also envisions a staged approach instead of undertaking the mammoth task of setting up a full-fledged DUNE detector of 40 kt fiducial mass~\cite{DUNE:2020jqi}. Therefore, it becomes imperative to discuss sensitivity study as a function of exposure. In figure~\ref{fig:Exposure_DM}, we study the nature of $\Delta \chi^2_{\rm DM}$ at 3$\sigma$ and 5$\sigma$ in Eq.~\ref{eq:deviation-from-maximality-chi2} as a function of scaled exposure for the standalone DUNE, T2HK, and their combination. We observe that initially, with an increase in exposure, the sensitivity to establish the deviation from maximal $\theta_{23}$ increases. However, the sensitivity after reaching half of their individual benchmark exposures reaches almost saturation. Horizontal illustrative dashed orange lines are drawn to depict the values of true $\sin^2\theta_{23}$ beyond which T2HK cannot differentiate between MM and the true values of $\sin^{2}\theta_{23}$ at its standard exposure, while the blue dots represent the intersection of the dashed orange line with the projected sensitivity curve of DUNE + T2HK. This implies that given the current benchmark choices of oscillation parameters in table~\ref{table:one}\,, the range of true values of $\sin^{2}\theta_{23}$ that can be differentiated from MM choices, by DUNE + T2HK with just $\sim 0.5$ of their nominal exposures, cannot be achieved by either of the individual experiments even with their respective projected exposures. When statistics are less, systematics become crucial. Therefore, at lower exposure, T2HK always performs better than DUNE to establish non-maximal $\sin^{2}\theta_{23}$, irrespective of the lower or higher octant of true choices of $\sin^{2}\theta_{23}$, because of better systematic uncertainties in disappearance rates. However, with the increment in exposure, the disappearance statistics in both DUNE and T2HK become similar. The complementary between DUNE + T2HK is essential to achieve a significant sensitivity at 5$\sigma$, even with high exposure. While the high precision measurements of DUNE in $\Delta m^{2}_{31}$ (due to substantial matter effect), complements the sensitivity of DUNE + T2HK at lower exposure, figure~\ref{fig:Exposure_DM} clearly shows that after a certain exposure, this study is no longer statistics-driven for achieving the sensitivity at 3$\sigma$. Nevertheless, a higher confidence level (5$\sigma$) is predominantly statistics-driven.

\subsection{Exclusion of wrong octant solutions of $\sin^{2}\theta_{23}$} 
\begin{figure}[htb!]
	\centering
	\includegraphics[width=0.8\linewidth]{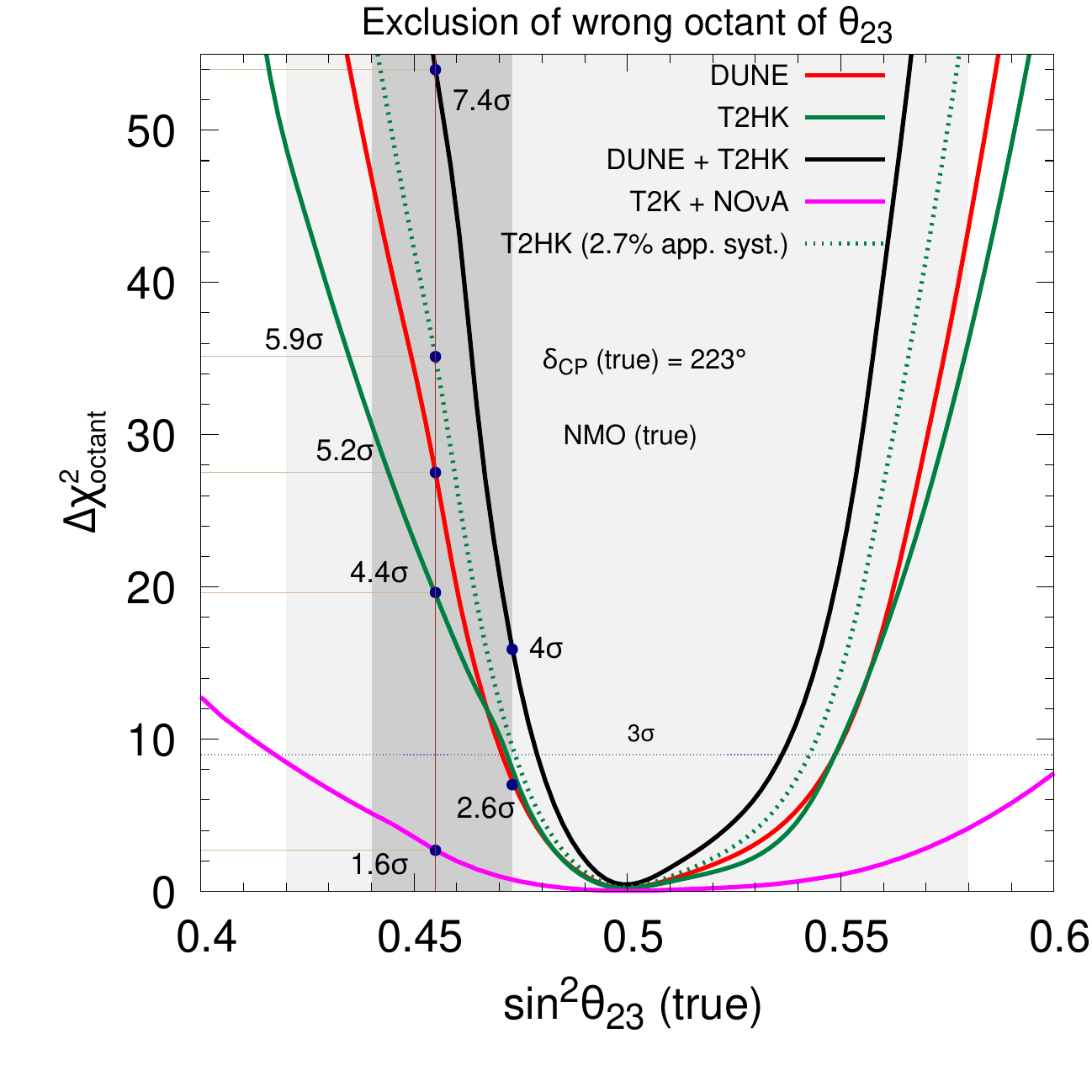}
	\caption{\footnotesize{ Sensitivity towards exclusion of wrong octant solutions in DUNE, T2HK, their combination, and T2K [2.5 yr $(\nu)$ + 2.5 yr $(\bar{\nu})$] + NO$\nu$A [3 yr $(\nu)$ + 3 yr $(\bar{\nu})$] as a function of $\sin^{2}\theta_{23}$ in data. The sensitivity of T2HK with an estimated improvement of nominal appearance systematic uncertainties to 2.7\% is also shown. In the fit, we marginalize over $\Delta m^{2}_{31}$ and $\delta_{\rm CP}$, keeping others fixed at their best-fit values (refer to table~\ref{table:one}). For illustrative purposes, we project out the statistical confidence for each curve, when $\sin^{2}\theta_{23} = 0.455$ in data.
 }}
	\label{fig:octant exclusion}
\end{figure}
\begin{figure}[htb!]
	\centering
	\includegraphics[width=\linewidth]{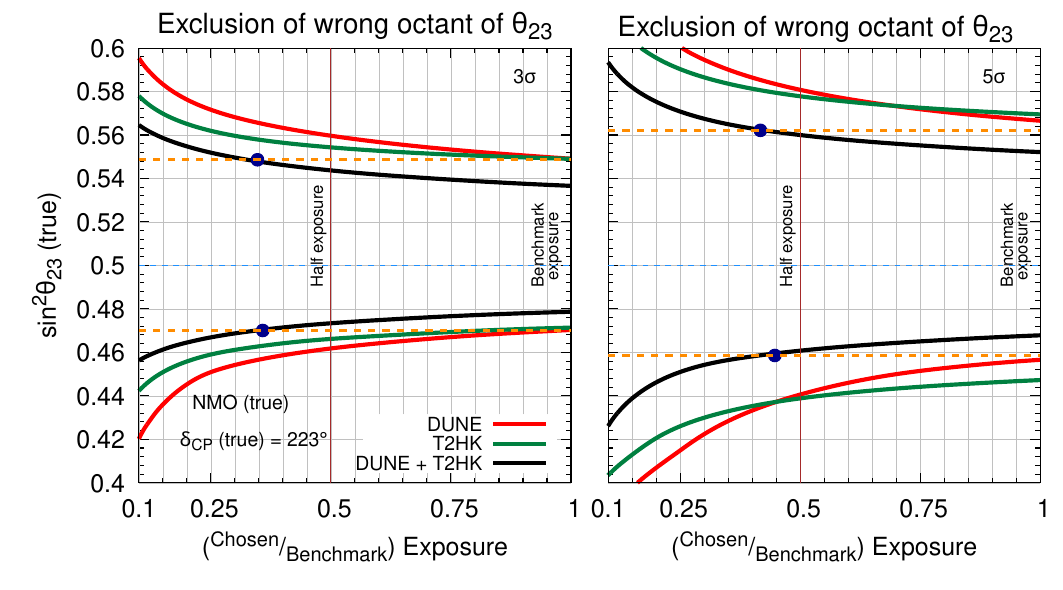}
	\mycaption{3$\sigma$  (5$\sigma$) exclusion of wrong octant of $\sin^{2}\theta_{23}$ as a function of scaled exposure is shown in left (right) panel. 1 depicts the benchmark exposure. We marginalize over the allowed regions of $\delta_{\mathrm{CP}}$ and $\Delta m^2_{31}$ in the fit. Refer to table~\ref{table:one}. \textit{ At lower exposure, the complementarity between DUNE + T2HK is the only solution to attain a 5$\sigma$ discovery in ruling out the wrong octant solutions for a significant range of $\sin^2 \theta_{23}$ in Nature. 
 } }
 \label{fig:octant-eclusion-vs-scaled-exposure}
 \end{figure}

Following the discussion of deviation from maximality, we study in this section the efficiency in establishing the octant of $\sin^{2}\theta_{23}$ by rejecting the hypothesis of wrong octant solutions. For this, we define 
\begin{eqnarray}
    \Delta \chi^{2}_{\text{octant}} &=& \underset{(\vec{\lambda})}{\mathrm{min}}\left\{ \chi^2\left(\sin^2\theta_{23}^{\mathrm{test}} \right) - \chi^2\left(\sin^2\theta_{23}^{\mathrm{true}}\right)\right\}\,,
\label{eq:octant-exclusion-chi2}
\end{eqnarray}
where $\sin^{2}\theta_{23}^{\mathrm{true}}$ refers to one octant, say LO; then, we generate data with [0.4, 0.5), while in the fit, we use the opposite octant, which in this case is HO $\in (0.5,0.6]$. Similar changes can be made by generating data with HO and excluding the LO hypothesis in the theory. In Eq.~\ref{eq:octant-exclusion-chi2}\,, $\vec{\lambda} = \{ \Delta m^{2}_{31}, \delta_{\rm CP}\}$. We use the corresponding allowed ranges from table~\ref{table:one}. Figure~\ref{fig:octant exclusion} depicts $\Delta \chi^{2}_{\text{octant}}$ as a function of $\sin^{2}\theta_{23}$ with which we generate data. It shows that alone DUNE and T2HK have similar sensitivity. T2HK is more stringent at lower significance, and DUNE is stronger at higher confidence. Large appearance systematic uncertainties in T2HK do not deteriorate the sensitivity, at least for lower significance due to comparable neutrino and antineutrino statistics~\cite{Agarwalla:2021bzs}. However, for attaining a higher significance, better systematic uncertainties in appearance rates are essential, which is a characteristic feature in DUNE (refer to section~\ref{sec:events}). The complementarity between DUNE and T2HK improves the standalone experiments' performance by almost $\sim 1.5$ times for the benchmark values from table~\ref{table:one}. The effect of improved appearance systematic uncertainties is distinctly visible when we consider the expected 2.7\% in T2HK~\cite{Munteanu:2022zla} instead of the nominal 5\%. Once the systematics are improved, T2HK performs better than DUNE irrespective of the true values of $\sin^{2}\theta_{23}$ in Nature. Sensitivity towards the exclusion of the wrong octant is dependent on both disappearance and appearance statistics, with the latter being dominant. For consistency, we have also checked the octant exclusion sensitivity using the best fit values and their corresponding allowed 3$\sigma$ ranges for minimization in the test-statistics from ref.~\cite{NuFIT:current}. We find that the results align closely with those shown in figure~\ref{fig:octant exclusion}.

In figure~\ref{fig:octant-eclusion-vs-scaled-exposure}\,, we study the efficacy of experiments in isolation and combination in ruling out the wrong octant of $\sin^{2}\theta_{23}$ as a function of scaled exposure. We observe that with 0.25 of their nominal exposures, DUNE alone will be able to differentiate about $\sim 45\%$ of $\sin^{2}\theta_{23}$ from wrong octant solutions; this improves to $\sim 50\%$ in T2HK, while their combination, DUNE + T2HK can differentiate $\sim 60\%$. So with just 0.25 of their individual exposures, it is possible in the combined setup to exclude the wrong octant for more than half of currently allowed $\sin^{2}\theta_{23}$ (refer to table~\ref{table:one}) at 3$\sigma$. Increasing beyond half of the nominal exposure does not help much, as the exclusion of the wrong octant solutions no longer remains statistics-driven. Further improvement in the allowed ranges of $\delta_{\mathrm{CP}}$ from the ongoing long-baseline experiments: T2K~\cite{T2K:2023smv} and NO$\nu$A~\cite{Carceller:2023kdz} may help to remove $(\sin^{2}\theta_{23} - \delta_{\mathrm{CP}})$ degeneracy and thus improve the sensitivity in wrong octant exclusion. We also notice that at 3$\sigma$ about $\sim 73\%$ of $\sin^{2}\theta_{23} \in [0.4, 0.6]$ can differentiate between correct and wrong octant solutions using the combined DUNE + T2HK setup, given the current benchmark values and projected exposure holds. For a higher confidence level (5$\sigma$), DUNE + T2HK is the only solution to attain sensitivity towards the exclusion of wrong octant solutions. We also observe that the discovery potential of DUNE + T2HK in excluding wrong octant solutions of $\sin^{2}\theta_{23}$, achievable with just $\sim$0.45 times of their individual exposures, is comparable to the sensitivity attained by standalone experiments using their nominal exposures.


\subsection{Precision measurements of $\sin^{2}\theta_{23}$ and $\Delta m^{2}_{31}$}
\label{precisionapp}

\begin{figure}[htb!]
\includegraphics[width=\linewidth]{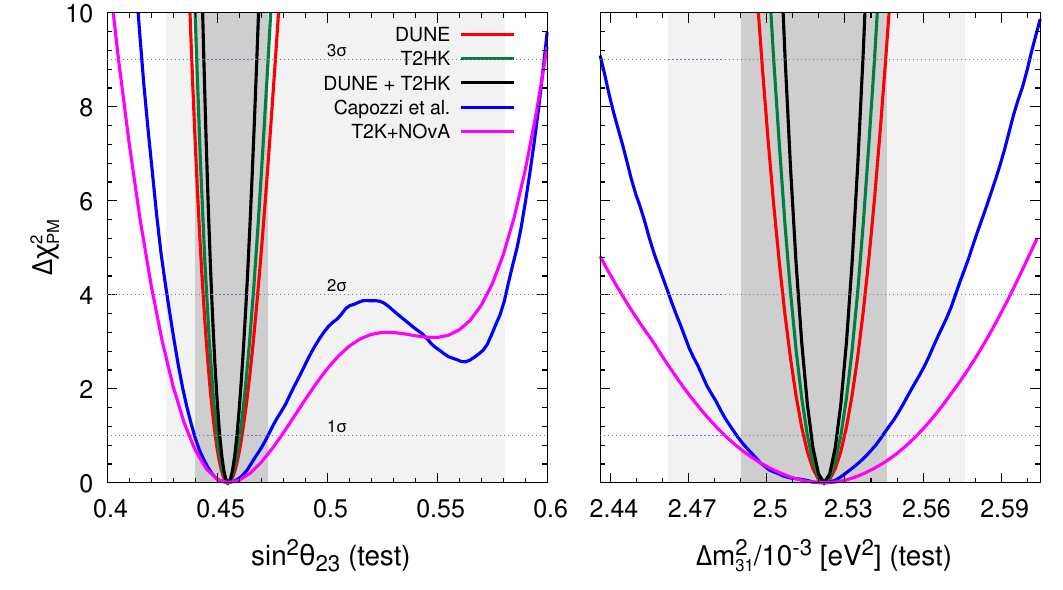}
\caption{\footnotesize{Expected achievable precision on $\sin^2\theta_{23}$ and $\Delta m^2_{31}$ around the respective benchmark values (refer to table~\ref{table:one}) using DUNE, T2HK, DUNE + T2HK, T2K [2.5 yr ($\nu$) + 2.5 yr ($\bar{\nu}$)] + NO$\nu$A [3 yr ($\nu$) + 3 yr ($\bar{\nu}$)], and global-fit from ref.~\cite{Capozzi:2021fjo}. In the fit, we marginalize over allowed region in $\delta_{\rm CP}$ and $\Delta m^{2}_{31}$ while determining precision on $\sin^2\theta_{23}$. Similarly, we perform marginalization over $\delta_{\rm CP}$ and $\sin^2\theta_{23}$ while determining precision measurements in $\Delta m^{2}_{31}$. Refer to table~\ref{table:four} for the computed values of the relative 1$\sigma$ precision, following Eq.~\ref{precision1}.  
}}
\label{fig:6}
\end{figure}
\begin{table}[t!]
	\centering
	\resizebox{\columnwidth}{!}{%
		\begin{tabular}{|c|c|c|c|c|c|c|}
			\hline
			\multirow{3}{*}{Parameter} & \multicolumn{6}{c|}{Relative 1$\sigma$ precision (\%)}\\
			\cline{2-7}
			& T2HK& DUNE & T2HK+DUNE & T2K+NO$\nu$A &Capozzi $et~al.$  & JUNO \\
			\hline
			$\sin^{2}\theta_{23}$ & 1.18 & 1.40 & 0.88 & 7.10 & 6.72 & ---\\
			\hline
			$\Delta m^{2}_{31}$ & 0.25& 0.31 & 0.20 & 0.99 & 1.09 & 0.2\\
			\hline
		\end{tabular}
	}
\caption{Relative 1$\sigma$ precision computed from figure~\ref{fig:6}, following Eq.~\ref{precision1}. In addition, we also give present-day global-fit precision from ref.~\cite{Capozzi:2021fjo} and expected relative 1$\sigma$ precision using JUNO with an estimated 6 years of run~\cite{NavasNicolas:2023fza}. }
\label{table:four}
\end{table}
\begin{figure}[t!]
	\centering
	\includegraphics[width=\linewidth]{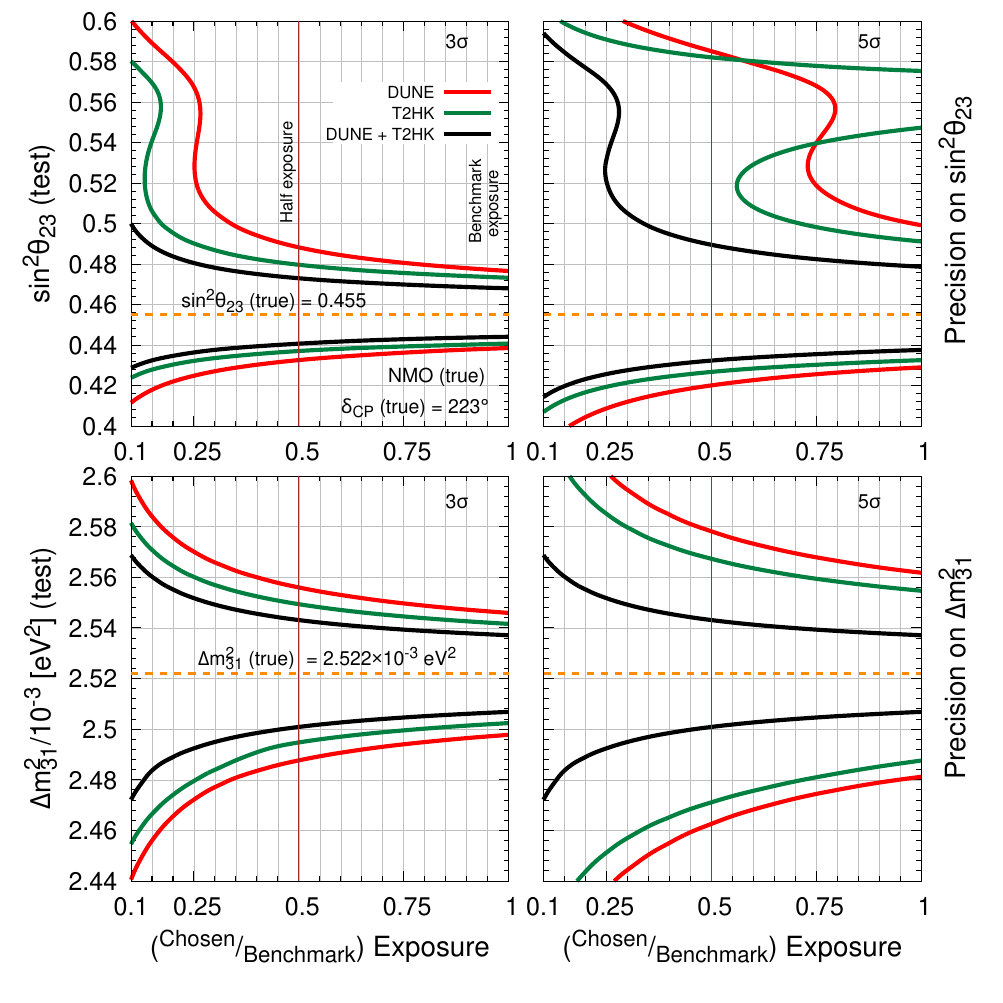}
	\mycaption{Precision on $\sin^2\theta_{23}$ and $\Delta m^2_{31}$ around their benchmark values (refer to table~\ref{table:one}) as a function of scaled exposure is shown. The upper (lower) panel depicts the precision on $\sin^2\theta_{23}$ ($\Delta m^2_{31}$) at two different C.L. In the fit, we marginalize over the allowed ranges in  $\delta_{\mathrm{CP}}$ and $\Delta m^2_{31}$ ($\sin^2\theta_{23}$) while producing upper (lower) panel. \textit{Considering the benchmark choices, high statistical confidence (5$\sigma$) precision on both $\sin^2\theta_{23}$ and $\Delta m^2_{31}$ that can be achieved by DUNE + T2HK with just 0.25 of their individual exposures cannot be attained by standalone experiments even with their full exposures.} }
	\label{figure:Exposure_precision}
\end{figure}
Following the exclusion of wrong octant solutions, it is imperative to question the precision in determining the value of atmospheric parameters: $\sin^2\theta_{23}$ and $\Delta m^2_{31}$. We compute the  statistical confidence for determining precision measurements on $\sin^{2}\theta_{23}$ by defining
\begin{equation}
\Delta \chi^2_{\text{PM}} = \underset{(\vec{\lambda})}{\mathrm{min}}\left\{ \chi^2\left(\sin^{2}\theta_{23}^{\mathrm{test}}\in [0.4,0.6] \right) - \chi^2\left(\sin^{2}\theta_{23}^{\mathrm{true}} = 0.455 \right)\right\}\,,
\label{precision-chisq-th23}
\end{equation}
while precision on $\Delta m_{31}^{2}$ is evaluated using
\begin{equation}
\Delta \chi^2_{\text{PM}} = \underset{(\vec{\lambda})}{\mathrm{min}}\left\{ \chi^2\left(\Delta m_{31}^{2,\mathrm{test}} \in [2.436,2.605]\times 10^{-3} \right) - \chi^2\left(\Delta m_{31}^{2,\mathrm{true}} = 2.522 \times 10^{-3} \right)\right\}\;.
\label{precision-chisq-m31}
\end{equation} 
 Here, test choices represent the corresponding allowed ranges of values, while the true choice is kept fixed at the benchmark choices (refer to table~\ref{table:one}). $\vec{\lambda}$ defines the set of oscillation parameters over which we perform marginalization in the fit given by $\vec{\lambda} = \{ \delta_{\rm CP}, \Delta m^{2}_{31}\}$ in Eq.~\ref{precision-chisq-th23} and $\vec{\lambda} = \{\delta_{\rm CP}, \sin^{2}\theta_{23}\}$ in Eq.~\ref{precision-chisq-m31}, respectively. For ease of quantifying, table~\ref{table:four} computes relative $1\sigma$-precision, defined as,
\begin{equation}
p(\zeta)\, =\, \frac{\zeta^{\rm max} - \zeta^{\rm min}}{6.0 \,\times \, \zeta^{\rm true}}\, \times \, 100\%\, .
\label{precision1}
\end{equation} 
Here, $\zeta^{\rm max}$ and $\zeta^{\rm min}$ depict the allowed upper and lower test values of each curve in the corresponding parameters (refer to figure~\ref{fig:6}) at $3\sigma$, respectively. We also quote the expected relative $1\sigma$-precision on $\Delta m^{2}_{31}$ from the upcoming reactor experiment, JUNO~\cite{NavasNicolas:2023fza} with the projected 6 years of runtime.

From figure~\ref{fig:6}\,, we observe that precision measurements in $\sin^2\theta_{23}$ allow a weak ($\sim 1.2\sigma$) clone solution in higher octant with the present global fit of oscillation data~\cite{Capozzi:2021fjo}. Although, with the full exposures of T2K + NO$\nu$A, we expect subtle improvement in it. However, the precision around the benchmark value of $\sin^2\theta_{23}$ is better using the present global fit oscillation data than the full projected exposures of present long-baseline experiments. This is because of huge disappearance statistics from other ongoing atmospheric experiments like Super-K and IceCube DeepCore. Similarly, we observe that the precision on $\Delta m^{2}_{31}$ using the global fit of oscillation data has already surpassed the expected precision using the full projected exposure of T2K + NO$\nu$A. This is mostly because of the input from reactor experiments like Daya Bay in the global fit. Additionally, we observe that due to the high energy resolution and large statistics in DUNE and T2HK, respectively, the standalone experiments can rule out the clone solution in $\sin^2\theta_{23}$. Comparatively, T2HK outperforms DUNE in precision measurements because of its extensive disappearance statistics and superior disappearance systematic uncertainties. A longer runtime in antineutrino mode also benefits T2HK, as both neutrino and antineutrino modes are crucial for achieving better precision measurements~\cite{Agarwalla:2013ju}. Table~\ref{table:four} helps in quantifying this improvement, showing the benefit of the interplay between DUNE and T2HK. Combining them improves the present-day~\cite{Capozzi:2021fjo} achievable precision on $\sin^2\theta_{23}$ and $\Delta m^2_{31}$ by a factor of $\sim$7 and $\sim$5, respectively. Further, we also make a comparison with the upcoming reactor experiment, JUNO~\cite{NavasNicolas:2023fza}. The achievable precision on $\sin^{2}\theta_{23}$ due to these next-generation experiments is truly remarkable. 

In figure~\ref{figure:Exposure_precision}\,, we study precision on both the atmospheric parameters as a function of scaled exposure. While the appearance channel is dominated by $\sin^{2}\theta_{23}$, disappearance channel is mostly influenced by $\sin^{2}2\theta_{23}$~\cite{Mikheyev:1985zog,Wolfenstein:1977ue,Cervera:2000kp,Freund:2001ui}. Hence, in resolving the issue of the wrong octant, appearance events play a crucial role, while for achieving better precision around the correct octant, disappearance events are essential. While standalone DUNE and T2HK bearing low exposures ($\sim 0.25$ times nominal exposure) cannot rule out clone solutions in $\sin^{2}\theta_{23}$ at 3$\sigma$, the combined DUNE + T2HK provide degeneracy-free measurements. Furthermore for achieving a discovery potential (5$\sigma$), standalone DUNE is unable to rule out the clone solutions in $\sin^{2}\theta_{23}$ even after achieving the projected exposure, T2HK needs $\sim 0.8$ of nominal exposure for excluding the wrong octant solutions. However, \textit{DUNE + T2HK can provide degeneracy-free precision on $\sin^{2}\theta_{23}$ at 5$\sigma$ by considering only $\sim 0.3$ times of their individual benchmark exposures}. In the standalone setup, DUNE performs better than T2HK because of the higher systematic uncertainties assumed in the appearance events of T2HK (5\% refer section~\ref{sec:2a}) than DUNE (2\%). As discussed earlier, appearance events are responsible for fixing the correct octant of $\sin^2\theta_{23}$ and thus removing any clone solutions. The combined precision displays the benefit of synergy between DUNE and T2HK, which can accomplish 5$\sigma$ precision around the correct octant. Furthermore, we observe that reaching 3$\sigma$ precision becomes saturated after a while; thus, it is no longer statistics-dominated. However, a degeneracy-free precision can be achieved by DUNE + T2HK at 5$\sigma$ even at lesser exposures if $\sin^{2}\theta_{23}$ turns out to be in LO in Nature. Similarly, in the case of $\Delta m^{2}_{31}$, \textit{using approximately 20\% of the individual exposures of DUNE and T2HK together can achieve an impressive relative 1$\sigma$ precision of 0.25\%. However, at the same exposure level, standalone DUNE and T2HK cannot distinguish between the benchmark values of $\Delta m^{2}_{31}$ and the currently allowed ranges from Table~\ref{table:one} when analyzed at 5$\sigma$.}

\section{Allowed regions in ($\sin^{2}\theta_{23}- \delta_{\mathrm{CP}}$) plane}
\label{sec:contour}
\begin{figure}[htb!]
	\centering
	\includegraphics[width=\linewidth]{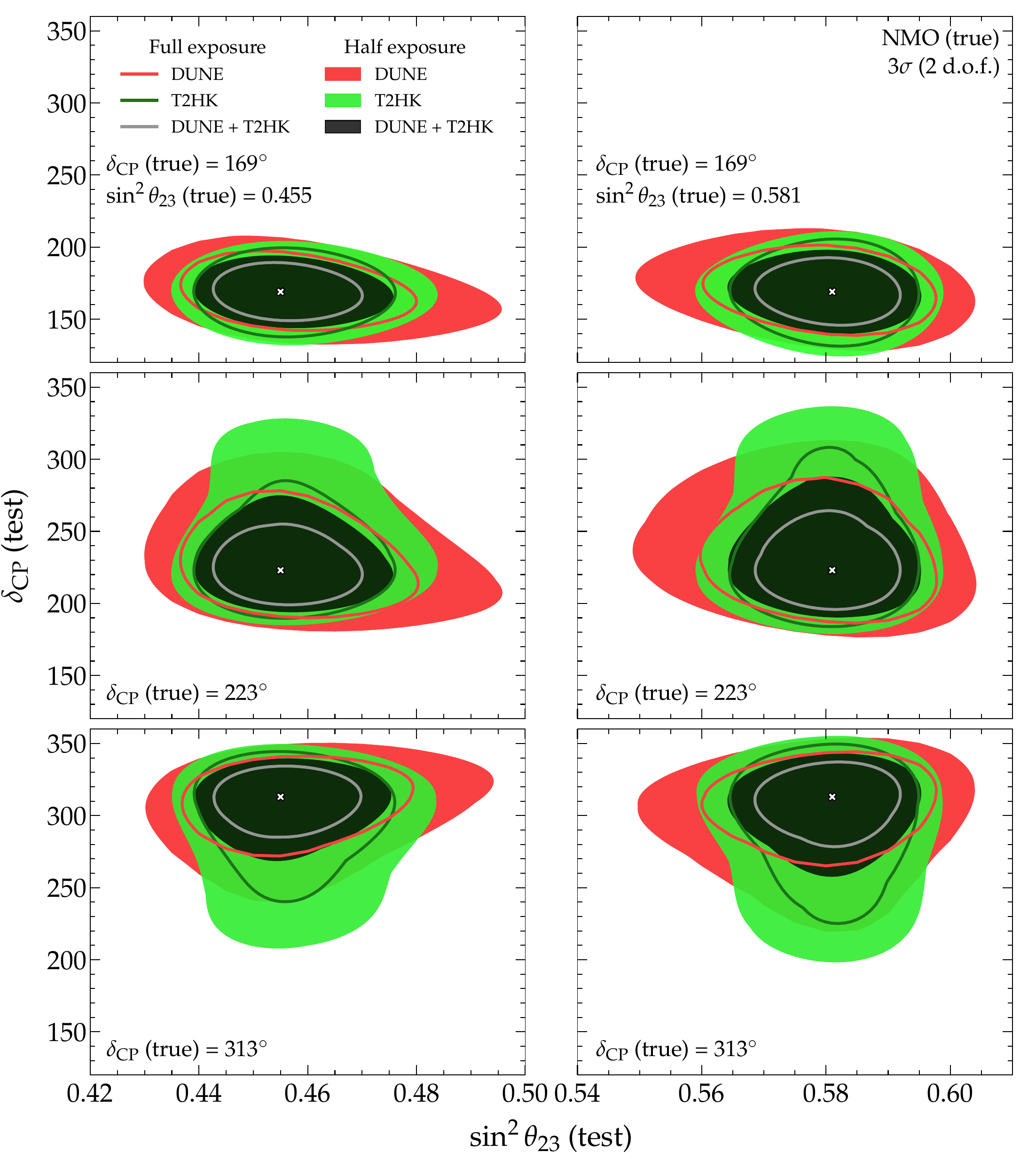}
	\mycaption{Allowed regions of the test atmospheric mixing angle, $\sin^2\theta_{23}$  and CP phase, $\delta_\mathrm{CP}$. The true values correspond to the benchmark values and other illustrative choices following ref.~\cite{Capozzi:2021fjo} (see section~\ref{sec:contour} for detail). The test-statistics is scanned over $\sin^{2}\theta_{23}$  and $\delta_{\mathrm{CP}}$ (refer to Eq.~\ref{eq:chi-allowed-ranges}). \textit{Combination is the only solution to exclude clone solutions hinted by the standalone experiments at half exposures.}
 }
	\label{fig:th23-dcp-allowed-ranges}
\end{figure}
As discussed previously, appearance events are necessary for extracting the correct octant of $\sin^2\theta_{23}$, while disappearance events are essential in obtaining a high-precision around the correct octant of $\sin^2\theta_{23}$. In this section therefore, we study the correlation between $\sin^{2}\theta_{23}$ and $ \delta_{\mathrm{CP}}$ in light of the current allowed oscillation parameter space. For this, we follow
\begin{eqnarray}
    \Delta \chi^{2} &=& \underset{(\vec{\lambda},\, \Delta m^{2}_{31})}{\mathrm{min}}\left\{ \chi^2\left(\sin^{2}\theta_{23}^{\mathrm{test}}\in [0.4,0.6], \delta_{\mathrm{CP}}^{\mathrm{test}} \in [0^{\circ}, 360^{\circ}] \right) - \chi^2 (\sin^{2}\theta_{23}^{\mathrm{true}},\delta_{\mathrm{CP}}^{\mathrm{true}} )\right\},
    \label{eq:chi-allowed-ranges}
\end{eqnarray}
where we use the benchmark values, $\pm 2\sigma \text{ in } \delta_{\mathrm{CP}} = 169^{\circ}$ and 313$^{\circ}$, and + 2$\sigma$ in $\sin^{2}\theta_{23} = 0.581$, following ref.~\cite{Capozzi:2021fjo} to generate data in each panel while scanning over the mentioned ranges in Eq.~\ref{eq:chi-allowed-ranges} of $\sin^2\theta_{23}$ and $\delta_{\mathrm{CP}}$. We marginalize over allowed ranges in $\Delta m^{2}_{31}$ (refer table~\ref{table:one}) in the fit.

In figure~\ref{fig:th23-dcp-allowed-ranges}\,, the benefit of exploiting the complementarity between DUNE and T2HK is clearly visible. DUNE with wide-beam is able to analyze various $L/E$ ratios. It also has access to the second oscillation maximum and 2\% appearance systematic uncertainties due to the magnificent LArTPC detector. These attribute to a better precision around the CP phase. Moreover, the large matter effect due to the long baseline helps in better precision measurement of $\Delta m^{2}_{31}$. However, more matter effect also induces extrinsic CP, which deteriorates the precision measurements in CP phase. This can be resolved by T2HK; with less matter effect, it provides better access to the intrinsic CP phase. Further, the huge disappearance statistics help in obtaining a better precision on $\sin^{2}\theta_{23}$. For comparison, the efficacy of DUNE + T2HK is visible in excluding the clone solutions hinted by the standalone experiments (in middle and lower panels), even with just half exposure. Further increasing the exposure from half to full in DUNE + T2HK leads to a quantitative decrement in the allowed region. For the upper panel, we observe that the allowed region is more or less similar in DUNE, T2HK, and the combination for the CP phase but varies with exposure for $\sin^{2}\theta_{23}$. This is because both DUNE and T2HK have better precision on $\delta_{\mathrm{CP}}$ when it is away from the CP-violating phases~\cite{Coloma:2012wq,Nath:2015kjg,Machado:2015vwa}. However, for achieving better precision on $\sin^{2}\theta_{23}$, large disappearance statistics are needed (therefore, T2HK consistently performs better than DUNE in this respect). Hence, a significant difference is visible when half exposure is increased to full exposure in the case of DUNE. However, the combination of both experiments is already performing well with their half exposures combined. While in the middle panel, we observe that the standalone experiments are hinting toward clone solutions --- T2HK in $\delta_{\mathrm{CP}}$ and DUNE in $\sin^{2}\theta_{23}$, the combination is able to precisely measure around the true values. The case in the lower panel is similar. Therefore, with half the exposures in each experiment, the combination is the only solution for better precision around the illustrative true values considered.

\section{Summary and conclusions}
\label{conclusion}
 The present generation of neutrino experiments have undoubtedly paved the way for precision studies in neutrino physics. The reactor mixing angle ($\theta_{13}$) measured by Daya Bay has achieved a remarkable precision of 2.8\%~\cite{DayaBay:2022orm}. Solar parameters: $\theta_{12}$ and $\Delta m^{2}_{21}$ have long been measured and stand presently with a relative 1$\sigma$ error of 4.5\% and 2.3\%\,, respectively from the global fit~\cite{Capozzi:2021fjo}. The most precisely measured oscillation parameter is, ironically, the magnitude of atmospheric mass-squared difference ($\Delta m^{2}_{31}$), while determining its sign still remains one of the big questions in neutrino physics left unanswered. The two most uncertain parameters are the atmospheric mixing angle, $\theta_{23}$, and Dirac CP phase, $\delta_{\mathrm{CP}}$. 

 In this work, we address the issue of maximal mixing solution of $\sin^{2}\theta_{23}$ and if non-maximal, then ruling out the wrong octant solutions of $\sin^{2}\theta_{23}$. We also study the achievable precision on $(\sin^{2}\theta_{23}-\delta_{\mathrm{CP}})$ and $(\sin^{2}\theta_{23}-\Delta m^{2}_{31})$ planes. To analyze the mentioned issues, we compare and contrast the standalone DUNE and T2HK with their complementarities in the combined DUNE + T2HK setup. While DUNE and T2HK, individually, should be able to improve on the sensitivity studies of deviation from maximal $\sin^{2}\theta_{23}$, exclusion of its wrong octant solutions, and precision measurements, their individual sensitivities are hampered by degeneracies due to uncertainties in $\sin^{2}\theta_{23}$, $\delta_{\mathrm{CP}}$, and $\Delta m^{2}_{31}$.  However, DUNE and T2HK have complementary capabilities: while T2HK is especially well-suited to measure $\sin^{2}\theta_{23}$ and $\delta_{\mathrm{CP}}$, DUNE is especially well-equipped to measure $\Delta m^{2}_{31}$. Thus, combining DUNE and T2HK brings many novelties.
 We find that the combined DUNE + T2HK increases the sensitivity to establish a deviation from maximality to $\sim 8\sigma$ if in Nature, $\sin^{2}\theta_{23}$ is same as the benchmark value mentioned in table~\ref{table:one}, following present global fit in ref.~\cite{Capozzi:2021fjo}. However, if this present best-fit shifts to its present $1\sigma$ allowed upper bound ($\sim 0.473$), then the combination is the only solution to achieve sensitivity to deviation from maximality greater than 3$\sigma$. The study of sensitivity towards deviation from maximal $\sin^{2}\theta_{23}$ as a function of exposure reveals that following the benchmark values, the discovery potential that combined DUNE + T2HK can reach with just 0.5 times the nominal exposure, is unattainable by either DUNE or T2HK even with their full exposures. In analyzing the sensitivity towards excluding incorrect octant solutions of $\sin^{2}\theta_{23}$, we find that the synergy of combined experiments significantly outperforms individual ones. While no single experiment achieves 5$\sigma$, the combination reaches approximately 8$\sigma$, assuming the true value of $\sin^{2}\theta_{23}$ is the same as the benchmark. Yet again, the discovery potential that standalone experiments attain with full exposure in eliminating the wrong octant, can be achieved by combining with just 0.4 times the full exposure. The estimated precision by alone DUNE and T2HK in both atmospheric parameters: $\sin^{2}\theta_{23}$ and $\Delta m^{2}_{31}$ improves present global fit precision by an approximate factor of 5 and 4, respectively. Furthermore, we find that the range of 5$\sigma$ precision that the combined DUNE + T2HK can achieve with only 0.25 times the exposure in $\Delta m^{2}_{31}$ is a level of precision that individual experiments cannot reach even with full exposures. Moreover, while studying the allowed ranges in $(\sin^{2}\theta_{23}- \delta_{\mathrm{CP}})$ plane, we notice that the weak hint towards clone solutions shown by standalone experiments at 3$\sigma$ can be resolved by the combination with just half the exposure. 
 
Therefore, this study brings about novel perspectives of the upcoming high-precision LBL experiments DUNE and T2HK, stressing how their combination and 
hence, the possible complementarities among them may alleviate the need of a very high exposure from these individual experiments in obtaining the desired
sensitivities towards different oscillation parameters.
 
 \subsection*{Acknowledgments}
We would like to thank F. Halzen, S. Prakash, and P. Swain for their useful comments on our work. S.K.A. would like to thank the organizers of the 16th International Conference on Heavy Quarks and Leptons (HQL 2023) at Tata Institute of Fundamental Research, Mumbai, Maharashtra, India, during 28th November to 2nd December 2023 and
the 29th International Workshop on Weak Interactions and Neutrinos (WIN 2023) at Sun Yat-sen University (SYSU) in Zhuhai , China, during 3rd to 8th July 2023, for 
providing an opportunity to present the preliminary results from this work. S.K.A. and R.K. acknowledge the support of the Department of Atomic Energy (DAE), Govt. of India, under the Project Identification no. RIO 4001. S.K.A. acknowledges the financial support from the Swarnajayanti Fellowship (sanction order no. DST/SJF/PSA- 05/2019-20) provided by the Department of Science and Technology (DST), Govt. of India, and the Research Grant (sanction order no. SB/SJF/2020-21/21) provided by the Science and Engineering Research Board (SERB), Govt. of India, under the Swarnajayanti Fellowship project. M.S. acknowledges the financial support from the DST, Govt. of India 
(DST/INSPIRE Fellowship/2018/IF180059). The numerical simulations are carried out using the ``SAMKHYA: High-Performance Computing Facility" at the Institute of Physics,, Bhubaneswar, India.

\begin{appendix}

\section{Allowed regions in ($\sin^{2}\theta_{23}-\Delta m_{31}^{2}$) plane}
\label{appendix1}

\begin{figure}[htb!]
	\centering
	\includegraphics[width=\linewidth]{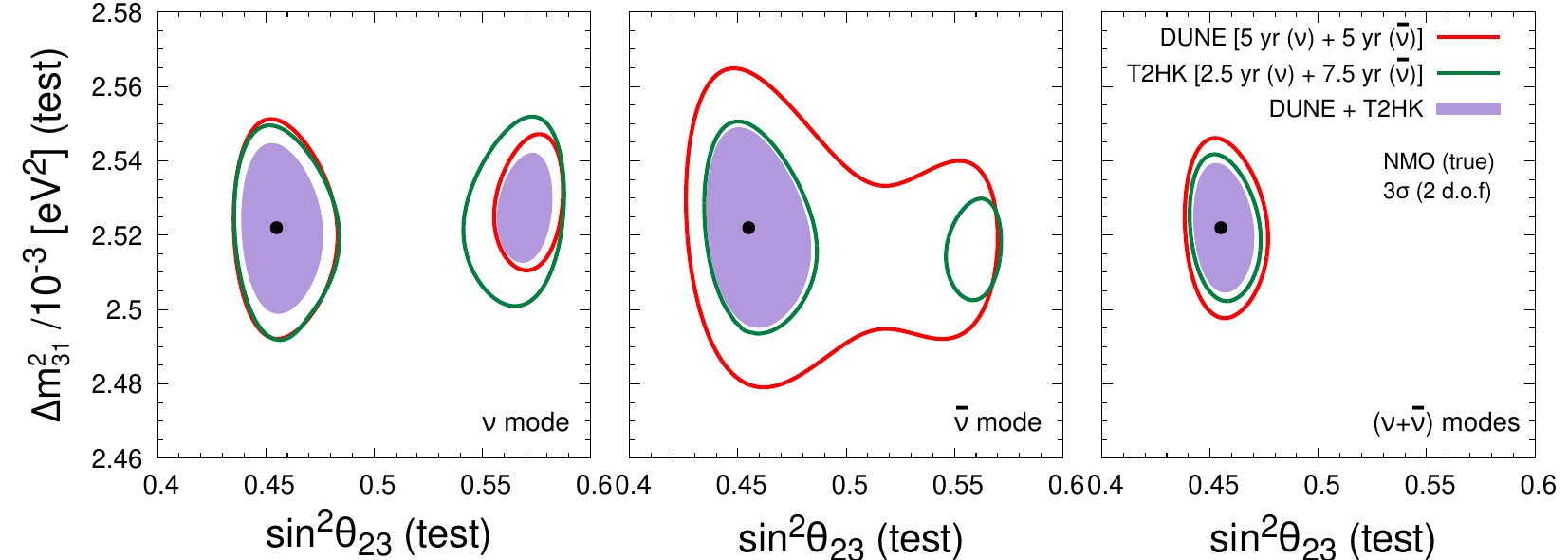}
	\mycaption{Allowed regions of the test atmospheric mixing angle, $\sin^2\theta_{23}$  and mass-squared splitting, $\Delta m^2_{31}$. The true values correspond to the benchmark values as mentioned in table~\ref{table:one}. The test-statistics is scanned over $\sin^{2}\theta_{23}$  and $\Delta m^2_{31}$ (refer Eq.~\ref{eq:chi-allowed-ranges-th23-Dm31}). We marginalize over the allowed ranges in $\sin^{2}\theta_{23}$ in the fit. \textit{In only antineutrino mode, DUNE + T2HK is the only solution to exclude the clone solutions in $\sin^2\theta_{23}$.}
}
        \label{fig:allowed-ranges-nmo}
\end{figure}

We follow
\begin{eqnarray}
    \Delta \chi^{2} &=& \underset{(\vec{\lambda},\, \sin^{2}\theta_{23})}{\mathrm{min}}\left\{ \chi^2\left(\sin^{2}\theta_{23}^{\mathrm{test}}, \Delta m_{31}^{2 ~ \mathrm{test}} \right) - \chi^2 (\sin^{2}\theta_{23}^{\mathrm{true}},\Delta m_{31}^{2 ~ \mathrm{true}} )\right\}\,,
    \label{eq:chi-allowed-ranges-th23-Dm31}
\end{eqnarray}
for generating figure~\ref{fig:allowed-ranges-nmo}, where true values correspond to the benchmark values as mentioned in table~\ref{table:one}, while scanning over the test-statistics for $\sin^{2}\theta_{23}\in [0.4,0.6]$ and allowed ranges of $\Delta m^{2}_{31}$ in table~\ref{table:one}. Further, we also perform marginalization over allowed ranges of $\sin^{2}\theta_{23}$ in the fit. We perform this study, separately for neutrino, antineutrino, and combined modes. 

As discussed previously with respect to DUNE~\cite{DUNE:2020jqi} and many other references in literature~\cite{Agarwalla:2013ju}\,, both neutrino and antineutrino modes are essential for breaking the $\sin^{2}\theta_{23} - \delta_{\mathrm{CP}}$ degeneracy and ruling out the wrong octant solutions in standalone experiments, while running in only antineutrino mode performs differently for the combined DUNE + T2HK. Alone, DUNE, when run in only antineutrino mode, is unable to expel even the MM solution of $\sin^{2}\theta_{23}$, T2HK has a clone solution at higher octant apart from the true octant. However, the complementary features in the combination are sufficient for breaking the $\sin^{2}\theta_{23} - \delta_{\mathrm{CP}}$ degeneracy in only antineutrino mode, which was not possible in standalone experiments. This is because, in the combined scenario, while T2HK has higher $\bar{\nu}$ statistics (refer to section~\ref{sec:2a}) leading to a majority of appearance events free of contamination from matter-induced CP phase, DUNE provides better precision measurement in $\Delta m^{2}_{31}$. The third panel shows the same curves as depicted in figure~\ref{fig:moneyplot} with full exposures. Isolated DUNE and T2HK are already able to achieve good precision around the best-fit breaking the $\sin^{2}\theta_{23} - \delta_{\mathrm{CP}}$ degeneracy, which gets more stringent around the best-fit with the combination.

\section{Sensitivity in ($\sin^{2}\theta_{23}-\Delta m_{31}^{2}$) plane with near-future experiments}
\label{appendix2}

\begin{figure}[htb!]
	\centering
	\includegraphics[width=\linewidth]{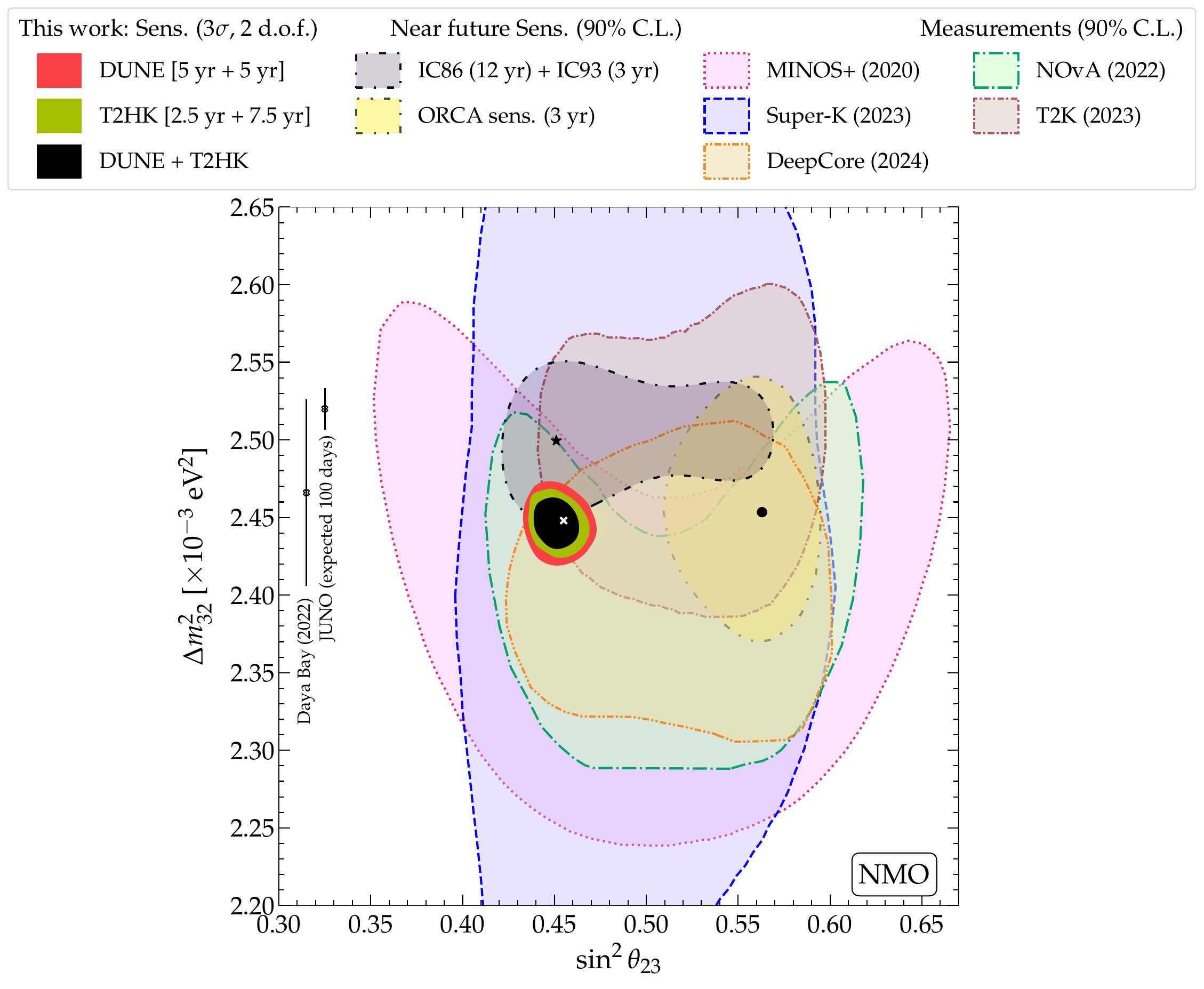}
	\mycaption{Allowed ranges at 3$\sigma$ (2 d.o.f.) in the atmospheric mixing parameters, $\sin^{2}\theta_{23}$ and $\Delta m^{2}_{32}$, using DUNE, T2HK, and DUNE + T2HK. DUNE is expected to have an exposure of 480 kt$\cdot$MW$\cdot$yr and T2HK, an exposure of 2431 kt$\cdot$MW$\cdot$yr. Same as figure~\ref{fig:moneyplot}, we show existing allowed ranges from: Super-K~\cite{linyan_wan_2022_6694761}, T2K~\cite{T2K:2023smv}, NO$\nu$A~\cite{NOvA:2021nfi}, MINOS~\cite{MINOS:2020llm}, and IceCube DeepCore~\cite{IceCube:2024xjj}. We also show existing (expected) bounds on $\Delta m^{2}_{32}$ from Daya Bay~\cite{DayaBay:2022orm} JUNO ~\cite{JUNO:2022mxj}). Additionally, we also show expected sensitivity from IceCube Deepcore (12 yr) + Upgrade (3 yr)~\cite{IceCube:2023ins}, and ORCA (3 yr)~\cite{KM3NeT:2021ozk}. Refer to the text for details.}
	\label{fig:appendix-allowed-ranges-nmo}
\end{figure}

In addition to the details in table~\ref{table:one}, in figure~\ref{fig:appendix-allowed-ranges-nmo}, we also show the expected allowed ranges using the near future development in the two giant atmospheric experiments: IceCube DeepCore and KM3NeT/ORCA. In ref.~\cite{IceCube:2023ins}, the expected sensitivity in atmospheric parameters is studied for 12 years of IceCube with 86 strings along with extra seven strings of IceCube Upgrade from 2026 onwards for three years. Strong improvements can be observed with the expected Upgrade in IceCube DeepCore. In ref.~\cite{KM3NeT:2021ozk}, the expected sensitivity of KM3NeT/ORCA after three years of data taking is studied, which we show in figure~\ref{fig:appendix-allowed-ranges-nmo}. The present scenario hints that the ongoing experiments in the near future will be able to obtain further precision on the $(\Delta m^2_{32} - \sin^{2}\theta_{23})$ plane. For comparison, we also show the assumed best-fit values by the two experiments: IceCube Upgrade ($\sin^{2}\theta_{23} = 0.451, \, \Delta m^2_{32} = 2.49 \times 10^{-3}$ eV$^{2}$) and ORCA ($\sin^{2}\theta_{23} = 0.563, \, \Delta m^2_{32} = 2.45 \times 10^{-3}$ eV$^{2}$). It should be noted that since their benchmark value for $\sin^{2}\theta_{23}$ corresponds to the opposite octant, the allowed region they expect seems complementary. 


\end{appendix}

\bibliographystyle{JHEP}
\bibliography{atmospheric}

\providecommand{\href}[2]{#2}\begingroup\raggedright\begin{thebibliography}{10}

\bibitem{DayaBay:2022orm}
{\bf Daya Bay} Collaboration, F.~P. An et~al., {\it {Precision Measurement of
  Reactor Antineutrino Oscillation at Kilometer-Scale Baselines by Daya Bay}},
  {\em Phys. Rev. Lett.} {\bf 130} (2023), no.~16 161802,
  [\href{http://arxiv.org/abs/2211.14988}{{\tt arXiv:2211.14988}}].

\bibitem{Super-Kamiokande:2023ahc}
{\bf Super-Kamiokande} Collaboration, T.~Wester et~al., {\it {Atmospheric
  neutrino oscillation analysis with neutron tagging and an expanded fiducial
  volume in Super-Kamiokande I\textendash{}V}},  {\em Phys. Rev. D} {\bf 109}
  (2024), no.~7 072014, [\href{http://arxiv.org/abs/2311.05105}{{\tt
  arXiv:2311.05105}}].

\bibitem{Ali:2022mrp}
{\bf T2K} Collaboration, A.~Ali, {\it {Precision Measurements of the PMNS
  Parameters with T2K Data}},  in {\em {20th Conference on Flavor Physics and
  CP Violation~}}, 7, 2022.
\newblock \href{http://arxiv.org/abs/2207.06496}{{\tt arXiv:2207.06496}}.

\bibitem{Catano-Mur:2022kyq}
{\bf NOvA} Collaboration, E.~Catano-Mur, {\it {Recent results from NOvA}},  in
  {\em {56th Rencontres de Moriond on Electroweak Interactions and Unified
  Theories}}, 6, 2022.
\newblock \href{http://arxiv.org/abs/2206.03542}{{\tt arXiv:2206.03542}}.

\bibitem{sk}
{\bf Super Kamiokande} Collaboration, Y.~Nakajima, {\it {Recent Results and
  future prospects from Super-Kamiokande}},  2020.
\newblock Talk given at the XXIX International Conference on Neutrino Physics
  and Astrophysics, Chicago, USA,
  \url{https://indico.fnal.gov/event/43209/contributions/187863/attachments/129474/159089/nakajima_Neutrino2020.pdf}.

\bibitem{IceCube:2019dyb}
{\bf IceCube} Collaboration, M.~G. Aartsen et~al., {\it {Development of an
  analysis to probe the neutrino mass ordering with atmospheric neutrinos using
  three years of IceCube DeepCore data}},  {\em Eur. Phys. J. C} {\bf 80}
  (2020), no.~1 9, [\href{http://arxiv.org/abs/1902.07771}{{\tt
  arXiv:1902.07771}}].

\bibitem{DayaBay:2018yms}
{\bf Daya Bay} Collaboration, D.~Adey et~al., {\it {Measurement of the Electron
  Antineutrino Oscillation with 1958 Days of Operation at Daya Bay}},  {\em
  Phys. Rev. Lett.} {\bf 121} (2018), no.~24 241805,
  [\href{http://arxiv.org/abs/1809.02261}{{\tt arXiv:1809.02261}}].

\bibitem{Seo:2019shs}
{\bf RENO} Collaboration, H.~Seo, {\it {Recent Result from RENO}},  {\em J.
  Phys. Conf. Ser.} {\bf 1216} (2019), no.~1 012003.

\bibitem{Capozzi:2021fjo}
F.~Capozzi, E.~Di~Valentino, E.~Lisi, A.~Marrone, A.~Melchiorri, and
  A.~Palazzo, {\it {Unfinished fabric of the three neutrino paradigm}},  {\em
  Phys. Rev. D} {\bf 104} (2021), no.~8 083031,
  [\href{http://arxiv.org/abs/2107.00532}{{\tt arXiv:2107.00532}}].

\bibitem{JUNO:2022mxj}
{\bf JUNO} Collaboration, A.~Abusleme et~al., {\it {Sub-percent precision
  measurement of neutrino oscillation parameters with JUNO}},  {\em Chin. Phys.
  C} {\bf 46} (2022), no.~12 123001,
  [\href{http://arxiv.org/abs/2204.13249}{{\tt arXiv:2204.13249}}].

\bibitem{IceCube:2024xjj}
{\bf IceCube} Collaboration, R.~Abbasi et~al., {\it {Measurement of atmospheric
  neutrino oscillation parameters using convolutional neural networks with 9.3
  years of data in IceCube DeepCore}},
  \href{http://arxiv.org/abs/2405.02163}{{\tt arXiv:2405.02163}}.

\bibitem{IceCube:2023ins}
{\bf IceCube} Collaboration, P.~Eller et~al., {\it {Sensitivity of the IceCube
  Upgrade to Atmospheric Neutrino Oscillations}},  {\em PoS} {\bf ICRC2023}
  (2023) 1036, [\href{http://arxiv.org/abs/2307.15295}{{\tt
  arXiv:2307.15295}}].

\bibitem{IceCube-Gen2:2019fet}
{\bf IceCube-Gen2} Collaboration, M.~G. Aartsen et~al., {\it {Combined
  sensitivity to the neutrino mass ordering with JUNO, the IceCube Upgrade, and
  PINGU}},  {\em Phys. Rev. D} {\bf 101} (2020), no.~3 032006,
  [\href{http://arxiv.org/abs/1911.06745}{{\tt arXiv:1911.06745}}].

\bibitem{KM3NeT:2021ozk}
{\bf KM3NeT} Collaboration, S.~Aiello et~al., {\it {Determining the neutrino
  mass ordering and oscillation parameters with KM3NeT/ORCA}},  {\em Eur. Phys.
  J. C} {\bf 82} (2022), no.~1 26, [\href{http://arxiv.org/abs/2103.09885}{{\tt
  arXiv:2103.09885}}].

\bibitem{KM3NeT:2024ecf}
{\bf KM3NeT} Collaboration, S.~Aiello et~al., {\it {Measurement of neutrino
  oscillation parameters with the first six detection units of KM3NeT/ORCA}},
  \href{http://arxiv.org/abs/2408.07015}{{\tt arXiv:2408.07015}}.

\bibitem{coelho_orca}
{\bf KM3NeT} Collaboration, J.~Coelho, {\it {Latest results from KM3NeT}},
  June, 2024.
\newblock Talk given at the XXXI International Conference on Neutrino Physics
  and Astrophysics, Milano,
  \url{https://agenda.infn.it/event/37867/contributions/233917/attachments/121916/178248/JCoelho_202406_Neutrino_KM3NeT.pdf}.

\bibitem{KM3NeT:2021rkn}
{\bf KM3NeT, JUNO} Collaboration, S.~Aiello et~al., {\it {Combined sensitivity
  of JUNO and KM3NeT/ORCA to the neutrino mass ordering}},  {\em JHEP} {\bf 03}
  (2022) 055, [\href{http://arxiv.org/abs/2108.06293}{{\tt arXiv:2108.06293}}].

\bibitem{T2K:2023smv}
{\bf T2K} Collaboration, K.~Abe et~al., {\it {Measurements of neutrino
  oscillation parameters from the T2K experiment using $3.6\times 10^{21}$
  protons on target}},  {\em Eur. Phys. J. C} {\bf 83} (2023), no.~9 782,
  [\href{http://arxiv.org/abs/2303.03222}{{\tt arXiv:2303.03222}}].

\bibitem{NOvA:2021nfi}
{\bf NOvA} Collaboration, M.~A. Acero et~al., {\it {Improved measurement of
  neutrino oscillation parameters by the NOvA experiment}},  {\em Phys. Rev. D}
  {\bf 106} (2022), no.~3 032004, [\href{http://arxiv.org/abs/2108.08219}{{\tt
  arXiv:2108.08219}}].

\bibitem{MINOS:2020llm}
{\bf MINOS+} Collaboration, P.~Adamson et~al., {\it {Precision Constraints for
  Three-Flavor Neutrino Oscillations from the Full MINOS+ and MINOS Dataset}},
  {\em Phys. Rev. Lett.} {\bf 125} (2020), no.~13 131802,
  [\href{http://arxiv.org/abs/2006.15208}{{\tt arXiv:2006.15208}}].

\bibitem{Yu:2023tmw}
{\bf IceCube} Collaboration, S.~Yu and J.~Micallef, {\it {Recent neutrino
  oscillation result with the IceCube experiment}},  in {\em {38th
  International Cosmic Ray Conference}}, 7, 2023.
\newblock \href{http://arxiv.org/abs/2307.15855}{{\tt arXiv:2307.15855}}.

\bibitem{deSalas:2020pgw}
P.~F. de~Salas, D.~V. Forero, S.~Gariazzo, P.~Mart\'\i{}nez-Mirav\'e, O.~Mena,
  C.~A. Ternes, M.~T\'ortola, and J.~W.~F. Valle, {\it {2020 global
  reassessment of the neutrino oscillation picture}},  {\em JHEP} {\bf 02}
  (2021) 071, [\href{http://arxiv.org/abs/2006.11237}{{\tt arXiv:2006.11237}}].

\bibitem{Esteban:2020cvm}
I.~Esteban, M.~C. Gonzalez-Garcia, M.~Maltoni, T.~Schwetz, and A.~Zhou, {\it
  {The fate of hints: updated global analysis of three-flavor neutrino
  oscillations}},  {\em JHEP} {\bf 09} (2020) 178,
  [\href{http://arxiv.org/abs/2007.14792}{{\tt arXiv:2007.14792}}].

\bibitem{NuFIT}
NuFIT v5.2 (2022), http://www.nu-fit.org/.

\bibitem{Fogli:1996pv}
G.~L. Fogli and E.~Lisi, {\it {Tests of three flavor mixing in long baseline
  neutrino oscillation experiments}},  {\em Phys. Rev. D} {\bf 54} (1996)
  3667--3670, [\href{http://arxiv.org/abs/hep-ph/9604415}{{\tt
  hep-ph/9604415}}].

\bibitem{Barger:2001yr}
V.~Barger, D.~Marfatia, and K.~Whisnant, {\it {Breaking eight-fold degeneracies
  in neutrino CP violation, mixing, and mass hierarchy}},  {\em Phys. Rev.}
  {\bf D65} (2002) 073023, [\href{http://arxiv.org/abs/hep-ph/0112119}{{\tt
  hep-ph/0112119}}].

\bibitem{Minakata:2002qi}
H.~Minakata, H.~Nunokawa, and S.~J. Parke, {\it {Parameter degeneracies in
  neutrino oscillation measurement of leptonic CP and T violation}},  {\em
  Phys.Rev.} {\bf D66} (2002) 093012,
  [\href{http://arxiv.org/abs/hep-ph/0208163}{{\tt hep-ph/0208163}}].

\bibitem{Minakata:2004pg}
H.~Minakata, M.~Sonoyama, and H.~Sugiyama, {\it {Determination of $\theta_{23}$
  in long-baseline neutrino oscillation experiments with three-flavor mixing
  effects}},  {\em Phys. Rev. D} {\bf 70} (2004) 113012,
  [\href{http://arxiv.org/abs/hep-ph/0406073}{{\tt hep-ph/0406073}}].

\bibitem{Hiraide:2006vh}
K.~Hiraide, H.~Minakata, T.~Nakaya, H.~Nunokawa, H.~Sugiyama, W.~J.~C. Teves,
  and R.~Z. Funchal, {\it {Resolving $\theta_{23}$ degeneracy by accelerator
  and reactor neutrino oscillation experiments}},  {\em Phys. Rev. D} {\bf 73}
  (2006) 093008, [\href{http://arxiv.org/abs/hep-ph/0601258}{{\tt
  hep-ph/0601258}}].

\bibitem{Agarwalla:2021bzs}
S.~K. Agarwalla, R.~Kundu, S.~Prakash, and M.~Singh, {\it {A close look on 2-3
  mixing angle with DUNE in light of current neutrino oscillation data}},  {\em
  JHEP} {\bf 03} (2022) 206, [\href{http://arxiv.org/abs/2111.11748}{{\tt
  arXiv:2111.11748}}].

\bibitem{DUNE:2020jqi}
{\bf DUNE} Collaboration, B.~Abi et~al., {\it {Long-baseline neutrino
  oscillation physics potential of the DUNE experiment}},  {\em Eur. Phys. J.
  C} {\bf 80} (2020), no.~10 978, [\href{http://arxiv.org/abs/2006.16043}{{\tt
  arXiv:2006.16043}}].

\bibitem{DUNE:2020ypp}
{\bf DUNE} Collaboration, B.~Abi et~al., {\it {Deep Underground Neutrino
  Experiment (DUNE), Far Detector Technical Design Report, Volume II: DUNE
  Physics}},  \href{http://arxiv.org/abs/2002.03005}{{\tt arXiv:2002.03005}}.

\bibitem{Hyper-KamiokandeProto-:2015xww}
{\bf Hyper-Kamiokande Proto-} Collaboration, K.~Abe et~al., {\it {Physics
  potential of a long-baseline neutrino oscillation experiment using a J-PARC
  neutrino beam and Hyper-Kamiokande}},  {\em PTEP} {\bf 2015} (2015) 053C02,
  [\href{http://arxiv.org/abs/1502.05199}{{\tt arXiv:1502.05199}}].

\bibitem{Hyper-Kamiokande:2018ofw}
{\bf Hyper-Kamiokande} Collaboration, K.~Abe et~al., {\it {Hyper-Kamiokande
  Design Report}},  \href{http://arxiv.org/abs/1805.04163}{{\tt
  arXiv:1805.04163}}.

\bibitem{T2K:2011qtm}
{\bf T2K} Collaboration, K.~Abe et~al., {\it {The T2K Experiment}},  {\em Nucl.
  Instrum. Meth. A} {\bf 659} (2011) 106--135,
  [\href{http://arxiv.org/abs/1106.1238}{{\tt arXiv:1106.1238}}].

\bibitem{NOvA:2007rmc}
{\bf NOvA} Collaboration, D.~S. Ayres et~al., {\it {The NOvA Technical Design
  Report}}, . FERMILAB-DESIGN-2007-01, 2007.

\bibitem{Huber:2004ka}
P.~Huber, M.~Lindner, and W.~Winter, {\it {Simulation of long-baseline neutrino
  oscillation experiments with GLoBES (General Long Baseline Experiment
  Simulator)}},  {\em Comput. Phys. Commun.} {\bf 167} (2005) 195,
  [\href{http://arxiv.org/abs/hep-ph/0407333}{{\tt hep-ph/0407333}}].

\bibitem{Huber:2007ji}
P.~Huber, J.~Kopp, M.~Lindner, M.~Rolinec, and W.~Winter, {\it {New features in
  the simulation of neutrino oscillation experiments with GLoBES 3.0: General
  Long Baseline Experiment Simulator}},  {\em Comput. Phys. Commun.} {\bf 177}
  (2007) 432--438, [\href{http://arxiv.org/abs/hep-ph/0701187}{{\tt
  hep-ph/0701187}}].

\bibitem{DUNE:2021tad}
{\bf DUNE} Collaboration, V.~Hewes et~al., {\it {Deep Underground Neutrino
  Experiment (DUNE) Near Detector Conceptual Design Report}},  {\em
  Instruments} {\bf 5} (2021), no.~4 31,
  [\href{http://arxiv.org/abs/2103.13910}{{\tt arXiv:2103.13910}}].

\bibitem{Hyper-Kamiokande:2016srs}
{\bf Hyper-Kamiokande} Collaboration, K.~Abe et~al., {\it {Physics potentials
  with the second Hyper-Kamiokande detector in Korea}},  {\em PTEP} {\bf 2018}
  (2018), no.~6 063C01, [\href{http://arxiv.org/abs/1611.06118}{{\tt
  arXiv:1611.06118}}].

\bibitem{DUNE:2021cuw}
{\bf DUNE} Collaboration, B.~Abi et~al., {\it {Experiment Simulation
  Configurations Approximating DUNE TDR}},
  \href{http://arxiv.org/abs/2103.04797}{{\tt arXiv:2103.04797}}.

\bibitem{Munteanu:2022zla}
L.-I. Munteanu, {\it {Long-baseline neutrino oscillation sensitivities with
  Hyper-Kamiokande}},  {\em PoS} {\bf NuFact2021} (2022) 056.

\bibitem{T2K:2014xyt}
{\bf T2K} Collaboration, K.~Abe et~al., {\it {Neutrino oscillation physics
  potential of the T2K experiment}},  {\em PTEP} {\bf 2015} (2015), no.~4
  043C01, [\href{http://arxiv.org/abs/1409.7469}{{\tt arXiv:1409.7469}}].

\bibitem{Patterson:2012zs}
{\bf NOvA} Collaboration, R.~Patterson, {\it {The NOvA Experiment: Status and
  Outlook}},  {\em Nucl.Phys.Proc.Suppl.} {\bf 235-236} (2013) 151--157,
  [\href{http://arxiv.org/abs/1209.0716}{{\tt arXiv:1209.0716}}].

\bibitem{Dore:2018ldz}
U.~Dore, P.~Loverre, and L.~Ludovici, {\it {History of accelerator neutrino
  beams}},  {\em Eur. Phys. J. H} {\bf 44} (2019), no.~4-5 271--305,
  [\href{http://arxiv.org/abs/1805.01373}{{\tt arXiv:1805.01373}}].

\bibitem{Katori:2016yel}
T.~Katori and M.~Martini, {\it {Neutrino\textendash{}nucleus cross sections for
  oscillation experiments}},  {\em J. Phys. G} {\bf 45} (2018), no.~1 013001,
  [\href{http://arxiv.org/abs/1611.07770}{{\tt arXiv:1611.07770}}].

\bibitem{Agarwalla:2022xdo}
S.~K. Agarwalla, S.~Das, A.~Giarnetti, D.~Meloni, and M.~Singh, {\it {Enhancing
  sensitivity to leptonic CP violation using complementarity among DUNE, T2HK,
  and T2HKK}},  {\em Eur. Phys. J. C} {\bf 83} (2023), no.~8 694,
  [\href{http://arxiv.org/abs/2211.10620}{{\tt arXiv:2211.10620}}].

\bibitem{Raut:2012dm}
S.~K. Raut, {\it {Effect of non-zero $\theta_{13}$ on the measurement of
  $\theta_{23}$}},  {\em Mod. Phys. Lett. A} {\bf 28} (2013) 1350093,
  [\href{http://arxiv.org/abs/1209.5658}{{\tt arXiv:1209.5658}}].

\bibitem{Baker:1983tu}
S.~Baker and R.~D. Cousins, {\it {Clarification of the Use of Chi Square and
  Likelihood Functions in Fits to Histograms}},  {\em Nucl. Instrum. Meth.}
  {\bf 221} (1984) 437--442.

\bibitem{Cowan:2010js}
G.~Cowan, K.~Cranmer, E.~Gross, and O.~Vitells, {\it {Asymptotic formulae for
  likelihood-based tests of new physics}},  {\em Eur. Phys. J. C} {\bf 71}
  (2011) 1554, [\href{http://arxiv.org/abs/1007.1727}{{\tt arXiv:1007.1727}}].
  [Erratum: Eur.Phys.J.C 73, 2501 (2013)].

\bibitem{Blennow:2013oma}
M.~Blennow, P.~Coloma, P.~Huber, and T.~Schwetz, {\it {Quantifying the
  sensitivity of oscillation experiments to the neutrino mass ordering}},  {\em
  JHEP} {\bf 03} (2014) 028, [\href{http://arxiv.org/abs/1311.1822}{{\tt
  arXiv:1311.1822}}].

\bibitem{Huber:2002mx}
P.~Huber, M.~Lindner, and W.~Winter, {\it {Superbeams versus neutrino
  factories}},  {\em Nucl. Phys.} {\bf B645} (2002) 3--48,
  [\href{http://arxiv.org/abs/hep-ph/0204352}{{\tt hep-ph/0204352}}].

\bibitem{Fogli:2002pt}
G.~L. Fogli, E.~Lisi, A.~Marrone, D.~Montanino, and A.~Palazzo, {\it {Getting
  the most from the statistical analysis of solar neutrino oscillations}},
  {\em Phys. Rev. D} {\bf 66} (2002) 053010,
  [\href{http://arxiv.org/abs/hep-ph/0206162}{{\tt hep-ph/0206162}}].

\bibitem{Gonzalez-Garcia:2004pka}
M.~C. Gonzalez-Garcia and M.~Maltoni, {\it {Atmospheric neutrino oscillations
  and new physics}},  {\em Phys. Rev. D} {\bf 70} (2004) 033010,
  [\href{http://arxiv.org/abs/hep-ph/0404085}{{\tt hep-ph/0404085}}].

\bibitem{Song:2020nfh}
N.~Song, S.~W. Li, C.~A. Arg\"uelles, M.~Bustamante, and A.~C. Vincent, {\it
  {The Future of High-Energy Astrophysical Neutrino Flavor Measurements}},
  {\em JCAP} {\bf 04} (2021) 054, [\href{http://arxiv.org/abs/2012.12893}{{\tt
  arXiv:2012.12893}}].

\bibitem{Ballett:2016daj}
P.~Ballett, S.~F. King, S.~Pascoli, N.~W. Prouse, and T.~Wang, {\it
  {Sensitivities and synergies of DUNE and T2HK}},  {\em Phys. Rev. D} {\bf 96}
  (2017), no.~3 033003, [\href{http://arxiv.org/abs/1612.07275}{{\tt
  arXiv:1612.07275}}].

\bibitem{deGouvea:2005hk}
A.~de~Gouvea, J.~Jenkins, and B.~Kayser, {\it {Neutrino mass hierarchy, vacuum
  oscillations, and vanishing |U(e3)|}},  {\em Phys. Rev. D} {\bf 71} (2005)
  113009, [\href{http://arxiv.org/abs/hep-ph/0503079}{{\tt hep-ph/0503079}}].

\bibitem{Coloma:2012ut}
P.~Coloma, T.~Li, and S.~Pascoli, {\it {A Comparative Study of Long-Baseline
  Superbeams within LAGUNA for large $\theta_{13}$}},
  \href{http://arxiv.org/abs/1206.4038}{{\tt arXiv:1206.4038}}.

\bibitem{NuFIT:current}
NuFIT v5.3 (2024), http://www.nu-fit.org/.

\bibitem{Carceller:2023kdz}
{\bf NOvA} Collaboration, J.~M. Carceller, {\it {3-flavour results with NOvA}},
   {\em PoS} {\bf NOW2022} (2023) 015.

\bibitem{NavasNicolas:2023fza}
{\bf JUNO} Collaboration, D.~Navas~Nicolas, {\it {Prospects of oscillation
  physics with JUNO}},  {\em PoS} {\bf NOW2022} (2023) 034.

\bibitem{Agarwalla:2013ju}
S.~K. Agarwalla, S.~Prakash, and S.~U. Sankar, {\it {Resolving the octant of
  $\theta_{23}$ with T2K and NOvA}},  {\em JHEP} {\bf 07} (2013) 131,
  [\href{http://arxiv.org/abs/1301.2574}{{\tt arXiv:1301.2574}}].

\bibitem{Mikheyev:1985zog}
S.~P. Mikheyev and A.~Y. Smirnov, {\it {Resonance Amplification of Oscillations
  in Matter and Spectroscopy of Solar Neutrinos}},  {\em Sov. J. Nucl. Phys.}
  {\bf 42} (1985) 913--917.

\bibitem{Wolfenstein:1977ue}
L.~Wolfenstein, {\it {Neutrino Oscillations in Matter}},  {\em Phys. Rev. D}
  {\bf 17} (1978) 2369--2374.

\bibitem{Cervera:2000kp}
A.~Cervera et~al., {\it {Golden measurements at a neutrino factory}},  {\em
  Nucl. Phys.} {\bf B579} (2000) 17--55,
  [\href{http://arxiv.org/abs/hep-ph/0002108}{{\tt hep-ph/0002108}}].
  [Erratum-ibid.B593:731-732,2001].

\bibitem{Freund:2001ui}
M.~Freund, P.~Huber, and M.~Lindner, {\it {Systematic exploration of the
  neutrino factory parameter space including errors and correlations}},  {\em
  Nucl. Phys. B} {\bf 615} (2001) 331--357,
  [\href{http://arxiv.org/abs/hep-ph/0105071}{{\tt hep-ph/0105071}}].

\bibitem{Coloma:2012wq}
P.~Coloma, A.~Donini, E.~Fernandez-Martinez, and P.~Hernandez, {\it {Precision
  on leptonic mixing parameters at future neutrino oscillation experiments}},
  {\em JHEP} {\bf 1206} (2012) 073, [\href{http://arxiv.org/abs/1203.5651}{{\tt
  arXiv:1203.5651}}].

\bibitem{Nath:2015kjg}
N.~Nath, M.~Ghosh, and S.~Goswami, {\it {The physics of antineutrinos in DUNE
  and determination of octant and $\delta_{CP}$}},  {\em Nucl. Phys. B} {\bf
  913} (2016) 381--404, [\href{http://arxiv.org/abs/1511.07496}{{\tt
  arXiv:1511.07496}}].

\bibitem{Machado:2015vwa}
P.~A.~N. Machado, {\it {Learning about the CP phase in the next 10 years}},
  {\em Nucl. Part. Phys. Proc.} {\bf 265-266} (2015) 174--176,
  [\href{http://arxiv.org/abs/1503.03775}{{\tt arXiv:1503.03775}}].

\bibitem{linyan_wan_2022_6694761}
{\bf Super-K} Collaboration, L.~Wan, {\it {Atmospheric Neutrino\_Super-K}},
  June, 2022.
\newblock Talk given at the XXX International Conference on Neutrino Physics
  and Astrophysics, Seoul, South Korea,
  \url{https://zenodo.org/record/6694761/files/Linyan%20Wan.pdf?download=1}.

\end{thebibliography}\endgroup

\end{document}